\documentclass{aastex}
\usepackage{emulateapj5}

\newcommand{\simgt}{\lower 2pt \hbox{$\, \buildrel {\scriptstyle >}\over {\scriptstyle\sim}\,$}}
\newcommand{\simlt}{\lower 2pt \hbox{$\, \buildrel {\scriptstyle <}\over {\scriptstyle\sim}\,$}}

\newcommand{\pgone}{PG~1115+080}
\newcommand{\apm}{APM~08279+5255}
\newcommand{\clover}{H~1413+117}

\newcommand{\chandra}{{\emph{Chandra}}}

\newcommand{\xmm}{\emph{XMM-Newton}}

\slugcomment{Received 2006 Oct 17; accepted 2007 Feb 27}
\shorttitle{Relativistic Fe Emission and Absorption in H 1413+117}
\shortauthors{CHARTAS ET AL.}

\begin{document}

\def\sarc{$^{\prime\prime}\!\!.$}
\def\arcsec{$^{\prime\prime}$}
\def\beginrefer{\section*{References}%
\begin{quotation}\mbox{}\par}
\def\refer#1\par{{\setlength{\parindent}{-\leftmargin}\indent#1\par}}
\def\endrefer{\end{quotation}}

\title{Discovery of Probable Relativistic Fe Emission and Absorption in the Cloverleaf Quasar 
H~1413+117}

\author{G. Chartas,\altaffilmark{1} 
M. Eracleous,\altaffilmark{1} \altaffilmark{2} \altaffilmark{3},
X. Dai,\altaffilmark{4} E. Agol,\altaffilmark{5} 
and S. Gallagher\altaffilmark{6}}

\altaffiltext{1}{Department of Astronomy \& Astrophysics, Pennsylvania State University,
University Park, PA 16802, chartas@astro.psu.edu}

\altaffiltext{2} {Center for Gravitational Wave Physics, The Pennsylvania State University, 104 Davey Lab, University Park, PA 16802}

\altaffiltext{3} {Department of Physics \& Astronomy, Northwestern University 2131 Tech Drive, Evanston, IL 60208}

\altaffiltext{4}{Department of Astronomy, The Ohio State University, Columbus, OH 43210}

\altaffiltext{5}{Astronomy Department, University of Washington, Box 351580, Seattle, WA 98195, USA}

\altaffiltext{6}{Department of Physics and Astronomy, UCLA, Mail Code 154705, 475 Portola Plaza, Los Angeles, CA 90095-1547}

\begin{abstract}

We present results from \chandra\ and \xmm\ observations of 
the low-ionization broad absorption line (LoBAL) quasar \clover.
Our spatial and spectral analysis of a recent deep \chandra\ observation confirms
a microlensing event in a previous \chandra\ observation performed about 5 years earlier.
We present constraints on the structure of the accretion flow 
in \clover\ based on the time-scale of this microlensing event.

Our analysis of the combined spectrum of all the images 
indicates the presence of two emission peaks at rest-frame 
energies of 5.35~keV and  6.32~keV
detected at the $ \simgt$  98\%  and $\simgt$ 99\% confidence levels, respectively.
The double peaked Fe emission line is fit well with an accretion-disk line model,
however, the best-fitting model parameters are neither well constrained nor unique.
Additional observations are required to constrain the 
model parameters better and to confirm the relativistic interpretation of the 
double peaked Fe K$\alpha$ line.
Another possible interpretation of the Fe emission 
is fluorescent Fe emission
from the back-side of the wind. 
The spectra of images C and D show significant high-energy broad absorption features
that extend up to rest-frame energies of 9~keV and 15~keV respectively.
We propose that a likely cause of these differences 
is significant variability of the outflow on time-scales that are shorter than the
time-delays between the images.

The \chandra\ observation of \clover\ has made possible for the first time the detection of the 
inner regions of the accretion disk and/or wind
and the high ionization component of the outflowing wind
of a LoBAL quasar.

\end{abstract}


\section{INTRODUCTION}

The broad X-ray absorption lines produced by outflowing material detected recently in two 
broad absorption line (BAL) quasars
possibly probe a highly ionized, high velocity component of the wind that 
appears to be distinct from the absorbers detected in the optical and UV wavebands 
(e.g., Chartas et al. 2007).
Because of the limited number of BAL quasars bright
enough to allow X-ray spectroscopy of their broad absorption lines,
the general X-ray properties of the outflows in such objects are poorly constrained.
The two X-ray brightest quasars with detected X-ray BALs
are APM 08279+5255 ($z$ = 3.91; Chartas et al., 2002)  and PG1115+080 ($z$ = 2.72; Chartas et al. 2003).
Order of magnitude estimates of the  mass outflow rates 
in these objects (Chartas et al. 2007) imply rates that are comparable to the estimated 
accretion rates of a few {M$_{\odot}~$yr$^{-1}$} and are considerably higher than 
those derived from observations of UV BALs.
The fraction of the total bolometric 
energy released by these two quasars into the IGM in
the form of kinetic energy (hereafter referred to as outflow efficiency) is constrained to be  
$\epsilon_{\rm k} =0.09_{-0.03}^{+0.07}$ , and
$\epsilon_{\rm k} = 0.3_{-0.1}^{+0.3}$, respectively. 
The main reason for the significantly higher outflow efficiencies
of the X-ray absorbing material compared to the UV absorbing material in these objects
is the significantly higher outflow velocities of the X-ray absorbing wind, reaching $\sim$0.4$c$
coupled with the fact that $\epsilon_{\rm k}$  $\propto$ $v_{wind}^{3}$.
Our derived estimates of the efficiency of the outflows in BAL quasars  \pgone\ and \apm,
when compared to values predicted by recent models of structure formation 
[e.g., Scannapieco \& Oh 2004 (SO04); Granato et al. 2004 (G04); 
Springel, Di Matteo, \& Hernquist 2005 (SDH05); Hopkins et al. 2005, 2006],
imply that these winds will have a significant impact on shaping the evolution of their host galaxies
and in regulating the growth of the central black hole. The outflowing X-ray absorbing material
is likely launched from the accretion disk. The high outflow velocities, the high ionization and the 
rapid flux variability of the X-ray BALs in \apm\ and \pgone\ indicate that the launching radius is 
close to the last stable orbit of the black hole (Chartas et al. 2002; 2003).
Fe~K$\alpha$ emission originating from the accretion disk was not detected in the
X-ray spectra of \apm\ and \pgone. This is not surprising since the
equivalent width of the Fe~K$\alpha$ line for the X-ray luminosities of these objects is 
expected to be $\simlt$ 200~eV (e.g., Nandra et al. 1997) which is below the detection limit of 
these observations. 
Part of the difficulty of detecting the Fe~K$\alpha$ line
in quasars arises from the significant dilution 
of the reprocessed emission from the disk by the emission from the primary X-ray source
(presumably the hot corona overlaying the inner disk that up-scatters the soft, 
thermal photons from the disk).
The equivalent width of the Fe line in quasars has also been observed to be inversely proportional
to luminosity (e.g., Nandra et al. 1997). One possible explanation is that the
disk density is reduced for more luminous quasars
and recent work indicates that it is proportional to 1/L (see Krolik 1998 eqn. 7.74).
The reduction of the disk density in luminous quasars
will result in a higher ionization parameter
in the surface layers of the disk causing a lower equivalent width for neutral iron line emission,
although leading to other features (e.g., Reynolds \& Nowak 2003).

Direct emission from the central X-ray source is thought to be significantly absorbed 
by an outflowing wind in a class of objects known as low ionization broad absorption line (LoBAL) quasars.
A recent {\it BeppoSAX} observation of the LoBAL quasar Mrk~231 (Braito et al. 2004)
confirmed the presence of a Compton thick absorber in this object with a column density of 
$N_{\rm H}$ $\sim$ 2 $\times$ 10$^{24}$~cm$^{-2}$.
Another member of this rare class is the gravitationally lensed $z$ = 2.56 quasar \clover\ (the Cloverleaf quasar).
A recent 38~ks \chandra\ observation of \clover\
taken in 2000 April revealed a remarkable iron emission feature 
(Oshima et al. 2001; Gallagher et al. 2002), which was
interpreted as Fe~K$\alpha$ fluorescence 
from the far side of the quasar outflow.
Our re-analysis of the Cloverleaf data (Chartas et al. 2004) 
indicated that the Fe~K$\alpha$ line 
was only significant in the brighter image A. 
We also found that the Fe~K$\alpha$ line and the continuum were
enhanced by different factors.
A microlensing event could explain both
the energy-dependent magnification and the significant detection of Fe~K$\alpha$
line emission in the spectrum of image A only. 
In the context of this interpretation we provided constraints on the
spatial extent of the inferred scattered continuum and reprocessed 
Fe~K$\alpha$ line emission regions in a  low-ionization broad absorption 
line quasar. From our analysis of the 38~ks \chandra\ observation of H~1413+117
we predicted that a follow-up observation would show the following, 
if microlensing was affecting image A:

(a) A reduction of the magnification of the X-ray continuum flux of image A
as the microlensing caustic traverses the accretion disk.
Once the microlensing event is complete the X-ray flux ratios of the images should
become similar to the $HST$ optical flux ratios in the $H$ and $R$  bands,
which are less sensitive to microlensing due to the larger sizescales of the $H$ and $R$  emission
regions compared to the X-ray emission region.

(b) A reduction in the Fe~K$\alpha$ line flux as the microlensing caustic sweeps 
across the accretion disk and outflowing wind.

In this work we present the results from a deeper 89~ks follow-up observation 
of H~1413+117 and revisit our microlensing interpretation
of the magnification event in image A. 
The deeper observation allows us to analyze of the spectra of individual images.
The spectra show broad absorption features bluewards of the rest-frame energy of $\sim$ 6.4~keV
and two remarkable emission line features redwards of this energy.
In the discussion section we present plausible interpretations of the properties
of the emission and absorption features.
We have also analyzed two \xmm\ observations of \clover\ taken 
between the two \chandra\ ones.
The \xmm\ observations do not resolve the lensed images but allow us to 
monitor the variability of the combined spectrum of all images of \clover.
Throughout this paper we adopt a $\Lambda$-dominated cosmology with 
$H_{0}$ = 70~km~s$^{-1}$~Mpc$^{-1}$, 
$\Omega_{\rm \Lambda}$ = 0.7, and  $\Omega_{\rm M}$ = 0.3.
The luminosity distance to \clover\ in this cosmology is $\sim$ 21~Gpc.
The luminosity distance estimate is taken from 
Caroll, Press, \& Turner (1992) and is based on 
the Friedman-Robertson-Walker model.

\section{OBSERVATIONS AND DATA ANALYSIS}

\clover\ was observed with the Advanced CCD imaging 
Spectrometer (ACIS; Garmire et al. 2003) on board the {\it Chandra X-ray Observatory} 
(hereafter \chandra) on 2000 April 19 and 2005 March 30 for 38.2~ks and 88.9~ks, respectively.
It was also observed with \xmm\ (Jansen et al. 2001) on 2001 July 29 and 2002 August 2 for 19.2~ks and 23.5~ks, respectively.
The spectral analysis of the \chandra\  April 2000 observation of \clover\ was presented in
Oshima et al. (2001), Gallagher et al. (2002), Chartas et al. (2004) and Dai et al. (2004).
Because of significant improvements in the calibration of the
instruments on board \chandra\ since the publication of these results, we have re-analyzed this observation.
Updates on the calibration of \chandra\ and \xmm\ are reported 
on the \chandra\ X-ray Center (CXC) and \xmm\ Science Operations Centre (SOC) 
World Wide Web (WWW) sites, respectively.
\footnote{The CXC and SOC WWW sites are located at 
\url{http://asc.harvard.edu/ciao/releasenotes/history.html } and \url{http://xmm.vilspa.esa.es/external/xmm\_sw\_cal/calib/rel\_notes/index.shtml}, respectively.}

The \chandra\ observations of \clover\ were analyzed using the standard 
software CIAO 3.3 provided by the CXC. 
For the reduction we used standard CXC threads to screen the data for 
status, grade, and time intervals of acceptable aspect solution and background levels.
The pointings of the observatory placed \clover\ on the back-illuminated S3 chip of ACIS.
To improve the spatial resolution we
removed a {$\pm$~0\sarc25} randomization applied to the event positions
in the CXC processing and employed a sub-pixel resolution technique
developed by Tsunemi et al. (2001) and later improved by Mori et al. (2001).

We analyzed the \xmm\ data for \clover\ with the standard analysis software SAS version 6.5 provided by the \xmm\ SOC.  We filtered the PN (Str{\" u}der et al. 2001) and MOS (Turner et al. 2001)
data by selecting events with \verb+PATTERNS+
in the 0--4 and 0--12 ranges, respectively.
Several moderate-amplitude background flares were present
during the \xmm\ observations of \clover.
The PN and MOS data were filtered to exclude times when the full-field count rates
exceeded 20~s$^{-1}$ and 4~s$^{-1}$, respectively.   

A log of the  \chandra\ and \xmm\ observations of \clover\ observations that includes 
observation dates, observed count rates, 
total exposure times, 
and observational identification numbers is presented in Table 1. 

In both the \chandra\ and \xmm\ analyses we tested the sensitivity of our results 
to the selected background and source-extraction
regions by varying the locations of the background regions and varying the 
sizes of the source-extraction regions. We did not find any significant change in the 
background-subtracted spectra. 
As we describe in \S2.2, \S2.3 and \S2.4 the \chandra\ and \xmm\ 
spectra of \clover\ were fitted with a variety of models
employing \verb+XSPEC+ version 12 (Arnaud 1996).
%For all models of \clover\ 
In all models
we included Galactic absorption due to neutral gas with a
column density of  $N_{\rm H}$=1.82 $\times$ 10$^{20}$~cm$^{-2}$ (Dickey \& Lockman, 1990). Falco et al. (1999) give a total extinction of 0.22 $\pm$ 0.04 mag for the lensing galaxy and Dai et al. (2006) estimate a dust-to-gas ratio of 
E($B-V$)/$N_{\rm H}$ = (1.4 $\pm$ 0.5) $\times$~10$^{-22}$~mag~cm$^{2}$ for the lens.  
Based on these results we estimate the column density of the lens 
to be $N_{\rm H}$ $\sim$ 1.5 $\times$~10$^{21}$.
We conclude that the absorption from the lens does not have any significant
effect on the spectral modeling of \clover\ since
the intrinsic absorption is significantly larger with a column density of 
$N_{\rm H}$ = 1--10~$\times$~10$^{23}$ (see \S2.2 and \S2.3).
All quoted errors are at the 90\% confidence level unless mentioned otherwise.
Due to the poorer S/N of the \xmm\ spectra of \clover\ all
errors on parameters obtained from fits to these spectra
are quoted at the 68\% confidence level.

\subsection{Spatial Analysis of the \chandra\ observation of \clover.}

To estimate the X-ray flux ratios of \clover\ we modeled the 89~ks \chandra\ images of 
A, B,  C and D, with 
point-spread functions (PSFs) generated by the simulation tool \verb+MARX+ (Wise et al. 1997).
The X-ray event locations were binned with a bin-size of 0\sarc0246 to sample the PSF 
sufficiently (an ACIS pixel subtends 0\sarc491).
The simulated PSFs were fitted to the \chandra\ data by minimizing the
$C$-statistic formed between the observed and simulated images
of \clover. 
In Figure 1a,b  we show the Lucy-Richardson deconvolved images in the 0.2--8~keV 
bandpass of the 38~ks and 89~ks \chandra\ observations, respectively.

We find that the X-ray flux ratios in the full (0.2-8keV) band for 
the 89~ks observation of \clover\ in March 2005 are 
[B/A]$_{full}$ = 1.02 $\pm$ 0.12, [C/A]$_{full}$ = 0.82$\pm$ 0.11, 
[D/A]$_{full}$ = 0.71 $\pm$ 0.09  and the number of detected X-ray events in
images A, B, C, and D were 160 $\pm$ 13, 163 $\pm$ 13, 131 $\pm$ 12, and 114 $\pm$ 11, respectively.
The X-ray flux ratios in the full (0.2-8keV) band for the 38~ks observation of \clover\ in April 2000 are 
[B/A]$_{full}$ = 0.49 $\pm$ 0.08, [C/A]$_{full}$ = 0.35$\pm$ 0.07, 
[D/A]$_{full}$ = 0.37 $\pm$ 0.07 and the number of detected X-ray events in
images A, B, C, and D were 147 $\pm$ 13, 72 $\pm$ 10, 52 $\pm$ 8, and 54 $\pm$ 9, respectively.

For comparison the {\sl HST} WFPC2 $F702W$-band flux ratios  
are [B/A]$_{F702W}$ = 0.87 $\pm$ 0.02, [C/A]$_{F702W}$ = 0.77 $\pm$ 0.01, 
[D/A]$_{F702W}$ = 0.72 $\pm$ 0.01. The HST WFPC2 $F702W$-band
observations were taken in 1994 December (Turnshek et al. 1997).
The HST WFPC2 $F702W$ band is centered at 6919 {\AA} 
with a bandwidth of 1385 {\AA} and represents a wide $R$-band.

We find that the X-ray flux ratios have varied significantly between 
the two \chandra\ observations. The X-ray flux ratios for the 2005 observation of 
\clover\ are consistent with the $F702W$-band flux ratios.  
The 0.2-8~keV count-rate of image A decreased by a factor of $\sim$ 2.1 
while the 0.2-8~keV count-rates of the other images remained the same within errors
between the 2000 and 2005 \chandra\ observations.

The full-band flux fractions of image A, [A/(B+C+D)], during the 38~ks and 89~ks \chandra\ observations
are  0.83 $\pm$ 0.10 and 0.39 $\pm$ 0.04.
For comparison the flux fraction of image A in the {\sl HST} WFPC2 $F702W$-band
is 0.42 $\pm$ 0.01.

\subsection{Spectral Analysis of the Individual Images of the \chandra\ observation of \clover.}

We performed fits to the spectra of individual images of \clover\
with a variety of models of increasing complexity.
We used events with observed-frame
energies lying within the range of 0.4--8~keV.
Due to the moderate S/N of the spectra we performed these fits using the Cash statistic
which does not require binning of the data. 
For comparison we also performed
the spectral fits using the $\chi^{2}$ statistic and found similar results. 
Our first model consisted of Galactic absorption and a simple power-law.
The best-fit parameters of this model are presented in Table 2 (model 1).
These fits are unacceptable in a statistical sense and the fit residuals show
significant absorption at observed-frame energies of $\sim$2--5~keV
and emission at observed-frame energies of $\sim$1--2~keV.

To illustrate better the possible presence of broad absorption features and emission lines we fitted
the spectra of the individual images of \clover\ in the observed-frame energy bands of
4.5--8~keV for image A, 5--8~keV for image B, 3--8~keV for image C and 4--8~keV for image D,
with simple power-law models modified
by Galactic absorption and extrapolated these models to lower energies.
The fits to the spectra of individual images and the residuals of these fits
are shown in Figure 2. We indicate the expected location of the
Fe K$\alpha$ emission line in these spectra and note that 
the possible broad absorption extends considerably above this energy.
We note that the fits to the high-energy range of the individual spectra 
are only used to qualitatively show the possible presence of absorption and emission
features. The significance of absorption and emission features in the individual spectra 
is estimated later in this section with fits to the observed-frame energy band of 0.4--8.~keV. 
The spectrum of image D shows an abrupt discontinuity in flux
near a rest-frame energy of $\sim$ 15~keV. 
Background contamination and instrumental effects are ruled out
as the origin of this feature for the following reasons:
(a) The background event rates in the 0.2--10. and 4.--6.~keV
bands in the spectrum of image D are 7.2 $\times$ 10$^{-6}$ and 4.5 $\times$ 10$^{-7}$, 
respectively compared to source event-rates 
of 1.1 $\times$ 10$^{-3}$ and 1.7 $\times$ 10$^{-4}$  in these bands.
Any background feature would also have to appear in all images
due to their close proximity.
(b) The ACIS response does not have any significant discontinuities in this
energy range and any instrumental effect would have to appear in all images
which is not the case.
A discontinuity in flux is also detected in the spectrum
of image C, near a rest-frame energy of $\sim$ 9~keV.
The residual plots also suggest possible Fe~K$\alpha$ emission lines 
in the spectrum of image C, near a rest-frame energy of $\sim$ 5.4~keV,
in the spectrum of image A at rest-frame energies of $\sim$ 5.4~keV and 
$\sim$ 6.25~keV and in the spectrum of 
image D at a rest-frame energy of $\sim$ 5.4~keV. 
These emission lines are more evident in the combined spectrum of all the images
shown in Figure 3.

As a refinement to our model we next included a neutral absorber at 
the redshift of the source (hereafter referred to as model 2 in Table 2).
The fits are significantly improved with the addition of an intrinsic neutral absorber 
with best-fit column densities ranging between 
9$_{-5}^{+7}$~$\times$~10$^{22}$~cm$^{-2}$ for image B and
29$_{-10}^{+21}$~$\times$~10$^{22}$~cm$^{-2}$ for image C (see model 2 in Table 2).
The best-fit spectral indices range between $\Gamma$ = 0.9$_{-0.4}^{+0.3}$ for image B
and $\Gamma$ = 1.9$_{-0.4}^{+0.7}$ for image C.

Later in the section we provide an 
estimate of the significance of the absorption and emission features 
in images C and D suggested in the residual plots of Figure 2.
Fits to the spectra of images C and D that model the high-energy 
absorption above the rest-frame energy of 6.4~keV with an absorption edge or 
single broad absorption line cannot successfully reproduce the abrupt apparent changes in flux at 
the rest-frame energies of $\sim$ 9~keV and $\sim$ 15~keV, respectively. 
We proceeded to model the spectra of images C and D with more complex models
motivated by our current understanding of the structure of 
LoBAL quasars (e.g., Green et al. 2001; Gallagher et al. 2002).

The next model considered (referred to as model 3 in Table 2) consists of Galactic 
absorption, a power-law continuum modified by intrinsic 
neutral absorption, a notch representing a saturated high-energy absorption line 
representing the X-ray BAL,
and a Gaussian emission line. The saturated absorption line model assumes absorption 
centered at energy $E_{\rm c}$ and within the energy range of 
$E_{\rm c}$ $\pm$ $E_{\rm w}$/2 the normalized intensity 
is equal to $1 - f_{\rm c}$, where $f_{\rm c}$ 
is the covering fraction.
\footnote{The covering factor, $f_{\rm c}$, effectively represents the fraction of 
photons from the background source(s) that pass through the absorber.}
 Fits to the spectra of images C and D using model 3 in 
Table 2 resulted in a significant improvement in fit quality 
compared to fits using model 2 in Table 2.
In particular, the $F$-test indicates that fits to the spectra of images C and D using model 3 in 
Table 2
resulted in an improvement of the fits compared to fits with model 2 in Table 2 at
the $\simgt$ 95\% and $\simgt $94\% confidence levels, respectively.

We also tried more sophisticated 
models (referred to as models 4 and 5 in Table 2) that consist of Galactic absorption, direct emission 
from a power-law modified by intrinsic 
neutral absorption, a saturated high-energy absorption line to account for the X-ray BALs,
scattered emission of a power-law assuming simple Thomson scattering 
to account for possible scattering of the central source emission from the outflowing wind,
and a fluorescent Fe~K$\alpha$ line from an accretion disk around the black hole.
We used Fe~K$\alpha$ line models that consider 
accretion onto a non-spinning (e.g., Fabian et al. 1989; model 4 of Table 2) and spinning black hole (e.g., Laor et al. 1991; model 5 of Table 2).
%The Fabian et al. and Laor et al.  models include general and special relativistic corrections %applicable to
%an accretion disk around a Schwarzschild
%and Kerr black hole, respectively.

Fits to the spectrum of image C using model 4 in Table 2 did not result in a significant  
improvement in fit quality compared to fits using model 2 in Table 2 (see Figure 4a).
Specifically, the $F$-test indicates a model that included an accretion-disk Fe~K$\alpha$ line at 
6.4~keV and a saturated absorption line centered at a rest-frame energy 
of 8.5 $\pm$ 0.2~keV(68\%) with rest-frame width of 
$E_{\rm w}$ = 0.5$_{-0.5}^{+0.3}$~keV(68\%)
resulted in a marginal improvement of the fit compared to fits with model 2 in Table 2 at the 82\% confidence level. The best-fit parameters are listed in Table 2.
Fits to the spectrum of image D using model 4 or model 5
that included an accretion-disk Fe~K$\alpha$ line at 
6.4~keV and a saturated absorption line centered at a rest-frame energy
of 14 $\pm$ 0.2~keV(68\%) with rest-frame width of 
$E_{\rm w}$ = 2.3 $\pm$ 0.4~keV(68\%) resulted 
in an improvement in fit quality at the 94\% confidence level 
compared to fits using model 2 (see Figure 4b).

\subsection{Spectral Analysis of the Combined Images of the \chandra\ observation of \clover.}

The spectra of the combined images of the 38~ks (2000 April) and 
89~ks (2005 March)  \chandra\ observations of \clover\ 
suggest the presence of emission line peaks redshifted with respect to the 6.4~keV energy of
Fe~K$\alpha$ line and absorption bluewards of this energy.
Our spectral analysis of the 89~ks \chandra\ observations of \clover\  indicated possible 
spectral differences between images. In particular, we detected significant 
differences between the maximum
absorption energies in images C and D. We therefore expect that the combined spectrum of the images 
to average out the high energy absorption structure found in the spectra of the individual image. 
With this caveat we fitted the combined spectra with models that included a power-law modified by intrinsic neutral absorption, two redshifted Gaussian emission lines and a broad Gaussian absorption line.
The spectra and best-fit models for the 38~ks and 89~ks 
observations are shown in Figures 3a and 3b respectively. 
These fits indicate the significant presence of 
emission line peaks at rest-frame energies of
E$_{\rm emis1}$ = 4.9$_{-0.4}^{+0.3}$~keV and 
E$_{\rm emis2}$ = 6.25 $\pm$ 0.25~keV 
for the 38~ks observation and
E$_{\rm emis1}$ = 5.35 $\pm$ 0.23~keV and 
E$_{\rm emis2}$ = 6.3$_{-0.3}^{+0.6}$~keV 
for the 89~ks observation. 
A conservative interpretation of these peaks 
based on their energies and the observational
fact that Fe~K$\alpha$ emission lines are common in quasars
is that they arise from Fe ~K$\alpha$ emission;
Recent spectral analyses (e.g., Porquet et al. 2004; Jimenez-Bailon et al. 2005; Schartel et al. 2005)
find significant detections of Fe K$\alpha$ emission lines in about half of the quasars
in their samples and there are indications of broadening of the lines in about 10\% of the quasars.

In Figures 5a and 5b we show the confidence contours between the
flux and rest-frame energy of the Fe~K$\alpha$ emission line peaks detected in the
38~ks and 89~ks observations of \clover.
All emission line peaks are detected at $ > $ 97\% confidence level
with the exception of the peak at 4.9~keV which is 
detected marginally at a confidence level of $\sim$ 85\%. 
We notice a statistically significant reduction in the flux of the blue Fe~K$\alpha$ line peak 
between these two observations. 
The red Fe~K$\alpha$ peak does not appear to vary within errors, however the detection of this peak in the 38 ks observation is marginal, with large error bars on the profile parameters.
%Because of the relative weakness of the red Fe~K$\alpha$ 
%line peak and its marginal detection in the 38~ks observation 
%we cannot place a tight constraint on its variability.
This decrease in the strength of the blue Fe~K$\alpha$ line peak is consistent with the 
microlensing interpretation proposed to explain
the anomalous flux ratios in the 38~ks observation of \clover. 
In particular, as a microlensing caustic traverses the Fe~K$\alpha$ emission line region 
we predicted the magnification of the Fe~K$\alpha$ line
to vary. The time separation between the two observations is about five years 
which is approximately the estimated duration of a microlensing event for this system.

We next fit the combined spectra with a more physically motivated model that consists of 
Galactic absorption, a power-law modified by intrinsic 
neutral absorption, a Gaussian absorption line,
and a fluorescent Fe~K$\alpha$ line from an accretion disk around a Schwarzschild black hole 
(model 2 in Table 3).
In Figure 6 we show the unfolded spectrum of \clover\ plotted with the best-fit model 
(model 2 in Table 3).
We also performed fits with a model 
similar to model 2 but with a line from a disk around
a Kerr black hole (model 3 in Table 3).
The fits using models 2 and 3 in Table 3 were acceptable in a statistical sense, 
however, because of the 
low S/N of the data we cannot constrain well the parameters
of the models and the solutions are not unique.

\subsection{Spectral Analysis of the \xmm\ Observations of \clover.}
{\it XMM-Newton} cannot spatially resolve the images of \clover\ and 
therefore we analyzed
the combined spectrum of all images. 

The spectral analysis of the 38~ks and 89~ks \chandra\ observations of \clover\ presented 
in Chartas et al. (2004) and in \S 2.2, respectively
indicated significant differences in the spectra of the different images. In particular,  
we detected significant enhancement of the Fe~K$\alpha$ line in image A during the 38~ks observation,
we found that the broad absorption features extended to different energies
blueward of the Fe~K$\alpha$ line, and inferred significant changes in the relative flux ratios between 
the two \chandra\ observations. 

Because of the significant spectral differences between the images, the
Hydrogen column densities, the Fe~K$\alpha$ emission line properties, and the broad absorption line properties
inferred from fits to the \xmm\ spectra of \clover\
represent average quantities from the four 
different spectra.

We first searched for broad absorption lines   
in the \xmm\ PN and MOS spectra by fitting 
them in the 4.5--10~keV band
%the spectra from 4.5--10~keV 
with a power-law model
(modified by Galactic absorption) and extrapolating this model to lower energies
(see Figure 7). The best-fit 4.5--10~keV spectral indices for the 2001 and 2002 observations
$\Gamma$ $\sim$ 1.4 and  $\Gamma$ $\sim$ 1 were not well constrained.
For the purpose of comparing the absorption residuals between the two \xmm\ observations of 
\clover\ we set the spectral index to 1.4.
The residuals of these fits in both the PN and combined MOS 1+2 detectors 
suggest absorption at rest-frame energies of 7--14~keV
and possible emission redward of the rest-frame energy of 6.4~keV,
however, the present analysis does not provide any 
useful constraints on the properties of the high energy absorber because of the 
low S/N of the \xmm\ data.

We next simultaneously fitted the PN, MOS1 and MOS2  
spectra of \clover\ at each of the  July 2001 and August 2002 observations 
with a model consisting of a power law modified by Galactic absorption and neutral intrinsic
absorption at $z = 2.56$ (model 1 in Table 4). 
These fits support the presence of an intrinsic absorber with
column densities of  $N_{\rm H}$= 25$_{-8}^{+9}$ $\times$ 10$^{22}$~cm$^{-2}$  (68\%),
and 19$_{-8}^{+11}$ $\times$ 10$^{22}$~cm$^{-2}$  (68\%)
the  2001 and 2002 \xmm\ observations of \clover\, respectively.

To obtain an estimate of the energies of possible Fe emission features we added 
a Gaussian emission line to our model. 
The best-fit rest-frame energies of these Gaussian lines
for the 2001 and 2002 observations are $5.8_{-2.5}^{+0.2}$~keV (68\%)
and $6.2_{-0.1}^{+0.3}$~keV (68\%), respectively. 
The results from the fits to the \xmm\ spectra of \clover\ are presented in Table 4.
For clarity we only show the higher S/N ratio PN data in Figure~7; however, all fits were performed simultaneously using the PN and MOS1+2.
The lower panels in Figure~7 show the $\Delta\chi^{2}$ residuals between the best-fit power-law model and the PN data.
Because of the low S/N of these spectra we did not attempt to fit these spectra with 
more complex models.

In \S 3.4 we use the X-ray fluxes obtained from the spectral analysis of the \xmm\ observations of \clover\ to infer the time evolution of the microlensing event in image A. 

\section{DISCUSSION}

Our analysis of the \chandra\ and \xmm\ observations
of \clover\ has confirmed the presence of a microlensing event in a previous 38~ks
\chandra\ observation and revealed the presence of
redshifted Fe~K$\alpha$ line peaks and broad absorption features
bluewards of Fe~K$\alpha$.
In $\S$ 3.1 and $\S$ 3.2 we investigate possible origins of these features
and provide constraints on the kinematic and physical properties of the medium 
producing them.
One of the surprising results of the spectral analysis was the 
detection of spectral differences between the images.
In $\S$ 3.3 we present plausible interpretations of these 
differences and suggest an observational 
strategy for discriminating between these interpretations.
In $\S$ 3.4  we revisit the microlensing interpretation 
of the magnification event in image A that we provided in Chartas et al. (2004) 
and infer constraints based on this interpretation.

\subsection{Origin of Emission Lines}

The emission features detected in the 89~ks \chandra\ observation
of \clover\ have energies near the Fe K$\alpha$ 6.4~keV line
and are reminiscent of the double peaked Fe~K$\alpha$ line profile with rest-frame energies of 5.3~keV and 6.4~keV
recently discovered in the radio-loud quasar PG 1425+267 with {\sl XMM-Newton} (Miniutti \& Fabian 2006).
These authors explored two possible origins of the line profile;
(a) a single relativistic iron line from the accretion disc
whose shape is determined by strong special and general relativistic 
effects, and 
(b) the superposition of a narrow 6.4~keV line from 
material at a large distance from the black hole
%distant material
and a relativistic one. 

We first consider the following scenario regarding the origin of the Fe~K$\alpha$ line peaks in \clover:
The double-peaked line arises from fluorescence from the far side of the accretion disk
and/or outflowing wind.
If we assume an outflow geometry for quasars similar to that proposed by
Proga et al. (2000) then, at sufficiently high inclination angles, the near side portion of the accretion disk is blocked
from our view by the outflowing X-ray absorbing material (see Figure 8).
The far side of the accretion disk, interior to the launching radius of the outflow,
is less absorbed and visible 
when the angle between our line of sight and the disk axis
is smaller than or equal
to the angle between the surface of the outflow and the disk axis. 
In the case of LoBAL quasars the high levels of UV polarization observed in several cases
suggest that our line of sight traverses a large portion 
of a Compton thick outflow and 
that the photons reaching the observer are mostly scattered and reprocessed.
\clover\ has a high level of polarization of $P$(5600Ð-6400 \AA) $\sim$ 2\% in the continuum
and a significant rise to $P$(4850Ð-4900 \AA) $\sim$ 4.5\% 
in the absorption troughs (Goodrich \& Miller, 1995).
Goodrich \& Miller propose that this difference in polarization levels between 
the continuum and the absorption troughs implies that the scattered light is less 
absorbed than emission from a direct line of sight.

Due to the relatively large observing beam-size
(for the ACIS  spatial resolution of $\sim$ 0\sarc5 the observing beam has a diameter of 
$\sim$ 4~kpc at the redshift of \clover)
we expect to detect emission from the far side of the accretion disk and the far side of the outflow.
This leads to the following question:
Are the observed properties of the emission line peaks in \clover\ (energy, energy width and equivalent width)
consistent with fluorescence from the far side of the accretion disk and/or the far side of the outflow?
Spectral analysis of broad Fe K$\alpha$ lines in several  Seyfert 1 galaxies  indicate that
the line emission originates between $\sim$ 6~$r_{\rm g}$ and $\sim$ 100~$r_{\rm g}$
(e.g., Nandra et al. 1997), where $r_{g} = GM_{bh}/c^{2}$ and $M_{bh}$ is the mass of the black hole.
The skewed and asymmetric structure of the Fe K$\alpha$ line observed in some Seyfert 1 galaxies 
is often interpreted as the result of special and general relativistic effects.
Based on spectral Model 1 of Table 3 we find the 2--10~keV luminosity 
of \clover\ to be
1.5 $\times$ 10$^{45}$/$\mu$ where $\mu$ is the lensing magnification and is expected to 
range between 20--40 (Chae et al. 1999).
Based on observations the expected EW of an Fe~K$\alpha$ line from the disk of an AGN as luminous as \clover\ is $\sim$ 250~eV (e.g., Nandra et al. 1997).
We note that the EW of the Fe~K${\alpha}$ line
is expected in theory to be a function of the inclination angle $i$ of
the accretion disk and the spectral slope $\Gamma$ of the incident
power-law spectrum as predicted in Monte Carlo calculations
of the interaction of X-rays with dense neutral material in standard disk models
(e.g., George \& Fabian 1991).
Based on our constraints on $\Gamma$ and $i$ from fits to the \chandra\ spectrum
of \clover\ these Monte Carlo  calculations predict the EW of the Fe~K$\alpha$ line in \clover\
to lie in the range of 150--200~eV for solar abundances (see Figure 14 of George \& Fabian 1991).
The rest-frame EWs of the Fe line peaks detected in \clover\ are poorly constrained 
to $\sim$500$_{-200}^{+1200}$~eV  and 
$\sim$1000$_{-900}^{+1200}$~eV, respectively.
In Table 3 we also list the EWs of the line peaks assuming disk line models.
These EWs are estimated with respect to the observed continuum.
The large equivalent widths of the detected line peaks can be explained
if emission from the central source is mostly absorbed.
A large fraction of the emission that we observe may be scattered emission from the wind or other 
scattering medium
and fluorescence from the far side of the accretion disk and/or outflow.
In Figure 9 we show the scattered emission spectrum, $f_{\rm scat}(E)$, assuming simple Thomson 
scattering and associated absorption of the continuum emission from a scatterer that subtends 
a solid angle of $\Omega_{\rm scat}$ to the central source. We have assumed 
that the photon flux density of the central source is a power-law of the form  
$f_{\rm cs}(E)$ $\propto$ $E^{-\Gamma}$, the spectral index is $\Gamma$ = 1.8, 
$\Omega_{\rm scat}$ = 0.15 and standard solar abundance values after 
Wilms, Allen, and McCray (2000).
In this simple model the estimated ratio of the incident emission spectrum
to that of the scattered emission spectrum 
at 6.4~keV is 
$f_{\rm cs}(E_{FeK\alpha})$/$f_{\rm scat}(E_{FeK\alpha})$  $\sim$ 18.

Based on the simple geometrical model shown in Figure 8 we estimated the EW of the 
Fe~K$\alpha$ line for the cases of
(a) no significant and (b) significant absorption from an outflowing wind. 
The $EW_{\rm noabs}$ of the Fe~K${\alpha}$ line for case (a) of no significant absorption from a wind has the form:

\begin{equation}
EW_{\rm noabs} \sim {{F_{\rm Fe}}\over{[f_{\rm disk}(E_{FeK\alpha}) + f_{\rm cor}(E_{FeK\alpha})]}}
\end{equation}

\noindent
where $F_{\rm Fe}$ is the integrated flux of the Fe~K${\alpha}$ line that 
originates from the disk,
$f_{\rm disk}(E_{FeK\alpha})$ and $f_{\rm cor}(E_{FeK\alpha})$ are the 
flux densities of the reprocessed continuum emission
from the disk and the corona respectively, estimated at the energy of the Fe line.
The $EW_{\rm abs}$ of the Fe~K${\alpha}$ line for case (b) where an 
absorber blocks the continuum from the corona and partially blocks the continuum from the disk has the form: 

\begin{equation}
EW_{abs} \sim {{A_{disk}F_{\rm Fe}}\over{[A_{disk}f_{\rm disk}(E_{FeK\alpha}) + A_{cor}f_{\rm cor}(E_{FeK\alpha})]}} \sim {{F_{\rm Fe}}\over{f_{\rm disk}(E_{FeK\alpha})}}
%EW_{abs} \sim {A_{1}f_{\rm Fe}}\over{(A_{1}f_{\rm disk} + A_{2}F_{\rm corona})} \sim {f_{\rm Fe}}\over{f_{\rm disk}}
\end{equation}

\noindent
where $A_{disk}$ and $A_{cor}$ are defined as the fractions of emission from the disk and corona that manage to penetrate the absorbing wind and be observed, respectively.
Since the reprocessed line and continuum are emitted
from the same area of the disk we assumed the same
fraction ($A_{disk}$) for line and continuum components. According to the geometrical model shown in Figure 8 we also assume
that the emission from the corona is completely blocked and therefore $A_{cor}$ $\sim$ 0.
The ratio of EWs with and without the Compton thick absorber is then given by the expression:

\begin{equation}
{{EW_{\rm abs}}\over{EW_{\rm noabs}}} \sim {1 + {{f_{\rm cor}(E_{FeK\alpha})}\over{f_{\rm disk}(E_{FeK\alpha})}}}
\end{equation}

Typically radio-quiet AGN have ${{f_{\rm cor}(E_{FeK\alpha})}/{f_{\rm disk}(E_{FeK\alpha})}}$ $\sim$ 10 
(see Figure 12 of George \& Fabian 1991).
We conclude that the large observed values of the EW of the line can be the result
of significant X-ray absorption of the direct emission from the central source. 
The detection, however, of possible X-ray BALs above 6.4~keV in images C and D
of \clover\ implies that a fraction of the direct emission apparently makes it through the 
absorbing wind at rest-frame energies above $\sim$ 6.4~keV. 

We also modeled the line profile in \clover\ assuming emission from a relativistic accretion disk
(e.g., Fabian et al. 1989; Laor et al. 1991). As shown in Figure 6 fitting the line with a disk model 
resulted in an acceptable fit to the spectrum. 
Because of the low S/N of the spectra the solutions obtained are not unique 
and the best-fit parameters are not well constrained.
In addition we caution that because of the spectral differences between images
the best-fit parameters should be considered as averaged quantities
between the four different spectra. The best-fit parameters are shown in Table 3.
In models 2 and 3 in Table 3 we consider the case of a Schwarzschild and Kerr
black hole, respectively. We note that the models that we used assume emission from the entire disk whereas we expect only the far side of the disk would be observed. Thus the 
observed line profiles would be 
quantitatively different from the model predictions but they would still be qualitatively similar.
To investigate the effects of disk obscuration on the Fe~K$\alpha$ line profile we
used the $KY$ relativistic accretion disk models developed by Dovciak et al. (2004).
We note that the low quality of the \chandra\ spectrum of \clover\ does not allow us
to uniquely constrain the parameters of the $KY$ models, however,
the goal of this exercise was to obtain a qualitative estimate of the
effects of obscuration on the shape of the Fe~K$\alpha$ emission line
and find a range of model parameters that are consistent with the data.
We fitted the \chandra\ spectrum of \clover\ with a model that
consists of Galactic absorption, a power-law modified by intrinsic
neutral absorption, a Gaussian absorption line,
and a fluorescent Fe~K$\alpha$ line from an accretion disk
around a black hole that is obscured over a range of azimuthal angles.
$kyr1line$ (Dovciak et al. 2004) was used to model the relativistic line from
an accretion disc around a Kerr black hole in the case of non-axisymmetric disk emission.
The inclination angle, and inner and outer radii of the disk
emission were fixed to the values of 30~degrees,  14~$r_{\rm g}$ and
17~$r_{\rm g}$, respectively, found from fits to the \chandra\ spectrum of \clover\ 
using the axisymmetric model $kyrline$.
In Figure 12 we show Fe~K$\alpha$ line profiles originating from five different azimuthal segments
of the disk assuming that portions of the disk are obscured by the outflowing
wind as illustrated in Figure 8. $\phi$ (units of degrees) is the lower azimuth of non-zero disk emissivity
and $\Delta\phi$ (units of degrees) is the span of the disk sector with non-zero disc emissivity.
Models with $-140$ degrees $ <  \phi <  -90 $ degrees, where substantially
more than half of the disk emission is obscured by the wind, are not consistent with the data.

We might also expect fluorescent Fe~K$\alpha$ line emission from the far side of the outflow. 
In this case the line profiles would also be quantitatively similar
to those computed above.
The Fe line emission detected in the 38~ks \chandra\
observation of \clover\ was initially interpreted by Oshima et al (2001) as Fe K$\alpha$ fluorescence 
from the far side of the outflow.
We note that in a re-analysis of the Cloverleaf observation (Chartas et al. 2004) we found
that interpretation to be problematic because of a significant microlensing event in image A.
The Fe~K$\alpha$ line in the 38~ks observation was only significant in image A 
and any constraints on the properties of the reprocessing medium based on
the strength of the Fe~K$\alpha$
line mentioned in that analysis are thus not reliable.

The present data cannot unambiguously constrain 
the origin of the Fe~K$\alpha$ emission. We briefly summarize and discuss the pros and cons of the two proposed interpretations :

(a)  Our first interpretation is that the observed line represents %emission originates from 
fluorescent Fe emission from the far side of the inner accretion disk.
The observed profile of the Fe~K$\alpha$ line is a result of 
special and general relativistic effects. Supporting evidence for this 
interpretation are the observed energies and widths of the line peaks.
The blue Fe~K$\alpha$ line peak at $\sim$ 6.3~keV is 
marginally broad (model 1 in Table 3)
which is consistent with an origin near the black hole. 
As we show in \S 3.4 the estimated size of the microlensed Fe emission region 
of the Fe peak at 6.3~keV is  $\sim$ 13~$r_{\rm g}$.
This is consistent with emission near the event horizon of the black hole.
The iron line peak at $\sim$ 5.4~keV is
significantly redshifted with respect to
6.4 keV implying also an origin near (10's of $r_{\rm g}$) the black hole.
In the context of this scenario, the line-emitting region of the disk must
be considerably larger than the source of the continuum (the corona) 
so that it is not completely obscured by the outflowing wind.
The extent, geometrical configuration and kinematics of the hot corona
are not well constrained from present observations and there are several competing models
for the geometry of the corona. Proposed configurations of the corona include
one where the cold accretion disk is surrounded by a hot extended, possibly patchy
corona (Haardt \& Maraschi 1991, Merloni 2003 ),
a point source located above the accretion disk 
(also referred to as the lamp-post model; e.g., Martocchia et al. 2002), 
multiple hot flares (Collin et al. 2003; Czerny et al 2004),
an aborted jet (Ghisellini et al. 2004), and a spherical central source spatially 
separate from the accretion disk (Proga 2005).
The lamp-post and central spherical corona models
would allow for different absorption towards the hot corona and the accretion disk. 

(b) Our second interpretation is that the Fe line emission is produced by 
fluorescence in the far side of the wind.
In this scenario the wind would have to be launched 
near the black hole and the fluorescent Fe emission would have to
originate from a narrow enough region to produce the 
double peaked structure and width of the observed iron line.

\subsection{Origin of Absorption Lines}

One plausible interpretation of the broad-absorption features detected bluewards of Fe K$\alpha$
(with significant detections in the spectra of images C and D)
is that they are resonance transitions in
highly ionized Fe XXV and/or Fe XXVI in an outflowing wind.
Among the most abundant elements of S, Si and Fe,  Fe absorption 
lines have the closest energies to the observed absorption features. 
Similar high-energy absorption features have been reported in 
BALQSOs \apm\ and \pgone\ (Chartas et al., 2002, 2003, 2007).
Assuming the above identification 
we infer maximum outflow velocities of 0.29$c$ and 0.67$c$ in the spectra of images
C and D respectively based on the maximum energy of the absorption troughs.
The rest frame widths of the absorption lines detected
in images C and D are $E_{\rm w}$ = 0.5$_{-0.5}^{+0.3}$~keV(68\%)
and $E_{\rm w}$ = 2.3 $\pm$ 0.4~keV(68\%).
One plausible interpretation of the observed broadening of the absorption lines 
is that it results from large velocity gradients 
in the outflow, along our line of sight.

In the following section we investigate plausible mechanisms to explain 
the difference in maximum absorption energies between images.

\subsection{Multiple Views of the Cloverleaf Quasar}

Three plausible interpretations for the spectral differences between images are:
(a) Significant inhomogeneities in the outflow properties along the lines of sight
corresponding to different images (e.g., Chelouche 2003; Green 2006),
(b) Variability of the outflow on time-scales that are shorter than the 
time-delay between the images, and
(c) Microlensing in at least two of the images.

{ \sl (a) Significant inhomogeneities in the outflow properties along the lines of sight
towards different images.}

We investigate this interpretation by estimating the linear separation
of the lines of sight as a function of the distance to the black hole
and compare these separations with the 
size of the central X-ray source.
The comoving linear beam separation is given by :

\begin{equation}
S(z) = \theta{{D_{ol}D_{sa}}\over{D_{sl}}}
\end{equation}

\noindent
where D are the angular diameter
distances, and the subscripts l, a, s, and o refer to the lens, absorber, source and observer.
In Figure 10a we show the linear beam separation as a function of redshift 
of the observer for an observed image separation of 1~arcsec assuming lens
and quasar redshifts of 1.55 and 2.56, respectively.
In Figure 10b we show the linear beam separation as a function of distance from the
black hole for a lens redshift of $z=1.55$.
The location of the material producing the 
high energy absorption is not constrained by the present observations.
We note however that similar high energy absorption
features have been detected in BAL quasars \apm\ and \pgone.
Based on the detected flux-variability of the X-ray BALs, the inferred outflow velocities
of the X-ray absorbing material, and the high ionization state of the
detected absorbers in these systems we inferred that they likely originate very close to
the black hole with inferred launching radii ranging between $\sim$ 10--100 $r_{\rm g}$.
For \clover\ we estimate 
$r_{\rm g}$ $\sim$ 4.5 $\times$ 10$^{14}$ $\mu$$^{-1}$ $\eta$$^{-1}$~cm 
$\sim$ 2.3 $\times$ 10$^{14}$~cm, where $\mu$ is the lensing magnification, 
$\eta$ = $L_{Bol}/L_{Edd}$, 
$L_{Edd}$ is the Eddington luminosity, $L_{Bol}$ is the bolometric luminosity and, $\eta$ is thought to range between 0.1 and 1 
in quasars. 
Our estimate of the dependence of the angular beam separation
with distance from the black hole shown in Figure 10b indicates 
that for outflowing absorbers at distances in the range of 
10--10$^{3}$ $r_{\rm g}$
the corresponding linear beam separations at these locations lie in the range of 
4.3$\times$10$^{10}$~cm -- 4.3$\times$10$^{12}$~cm.
These linear beam separations are significantly less than the size of the central X-ray source (a source size of 10$r_{g}$ $\sim$ 2.3 $\times$ 10$^{15}$~cm)
and therefore the \chandra\ beams (i.e., cylinders of sight) 
corresponding to the different images of \clover\ 
overlap almost completely, thus they sample the same regions of the outflow.
We conclude that the observed spectral differences between images cannot be caused by  
inhomogeneities in the outflow properties along different lines of sight.

{\sl (b) Variability of the outflow on time-scales that are shorter than the 
time-delays between the images.}

From a light travel time argument we estimate that the minimum 
variability time-scale in the observed-frame is 
$t_{\rm min}$ = 6(1 + $z$)$r_{\rm g}$/$c$ $\sim$ 2~days,
where 6$r_{\rm g}$ is the radius of the innermost stable circular
orbit around a Schwarzschild black hole.
We note that the last stable orbit can be even smaller in a Kerr black hole
and the disk can extend down to the event horizon.
The dynamical time-scale corresponding to motion near the event horizon across the line of sight 
is $\tau_{\rm dyn}$ = $\sqrt{(r/r_{\rm g})}$~$t_{\rm min}$ $\sim$ 5~days.
The dynamical time-scale $\tau_{\rm dyn}$ is perhaps more plausible than the light crossing-time $t_{\rm min}$  because
it corresponds to physical changes (e.g., motion across the line of sight or internal changes).
Variability of the outflow over time-scales that are shorter than the time-delays 
between images can result in spectral difference between images.

Lens models of \clover\ assuming that the lens consists of two galaxies (e.g., Chae et al. 1999) indicate that image C 
leads images B, A, and D by 15, 18.5 and 22 days, respectively. 
These values of the time-delays are sensitive to the adopted lens model,
however, the two-galaxy lens models and the galaxy $+$ cluster lens models considered by Chae et al. (1999)
both have C being the leading image and D being the trailing image.
We conclude that variability of the outflow can be the cause 
of the observed spectral differences since $t_{\rm dyn}$ is less than the time-delays.

We use the relative time-delays between images C and D to 
interpret the observed difference in the maximum absorption energies
in their spectra.
The spectrum of image C shows absorption of up to 9~keV in the rest-frame of \clover\
and the spectrum of image D shows absorption of up to 15~keV in the rest-frame.
The relative time-delays predicted in the Chae et al. (1999) model
imply that the spectrum of image D corresponds to an earlier rest-frame time than that of
the spectrum of image C. Assuming that the broad absorption is caused by X-ray absorbing material
in an outflow, one plausible interpretation of the observed differences in the maximum absorption energies between the spectra of 
imagers C and D is a decrease of the apparent velocity of the absorbing material
(i.e., the projection of its velocity along the line of sight).

The first detected change in the apparent velocity of an outflow 
associated with a Seyfert 1 galaxy was recently reported by Gabel et al. (2003) 
based on {\sl HST} monitoring observations of NGC 3783. 
They interpreted this velocity shift as either produced by a radial deceleration of a UV absorbing 
cloud or caused by the evolution of a continuous flow tube across our line of sight to the emission source. 
A similar mechanism may be responsible for the differences in maximum absorption energies
of the X-ray BALs in images C and D.

The variability interpretation is testable with a follow-up observation of \clover\ with
comparable exposure. In particular, if the spectral differences between 
images are due to intrinsic variability of the outflow over
time-scales shorter than the time-delay we predict 
a significant variability in the structure of the high-energy absorption
of the same image between \chandra\ observations in a future campaign.

{\sl  (c) Microlensing in at least two of the images.}
Our present spectral and spatial analysis of the 89~ks observation
of \clover\ indicate that the X-ray flux ratios have become similar to those 
observed in the optical $F702W$-band that is thought to be less sensitive to microlensing.
The analyses presented in \S 2.1 and \S 3.4 indicate that the microlensing event 
in image A inferred to have occurred during the 2000 observation of \clover\ has ended 
and there is no indication of any anomalous flux ratios in any images during the 2005 observation of \clover.
We conclude that microlensing is not the cause
of the spectral differences between the images in the 89~ks \chandra\ observation of \clover.

 \subsection{Revisiting the Microlensing Interpretation of the Magnification Event in Image A}

In our earlier analysis of the 38~ks \chandra\ observation of \clover\ (Chartas et al. 2004)
we determined that a microlensing event 
resulted 
in a significant enhancement of the Fe~K$\alpha$ line in image A
compared to that in the remaining images.
To estimate the evolution of the microlensing event in image A with time 
we calculate the observed-frame 1--2~keV photon flux of image A.
The 1--2~keV region contains the redshifted Fe K$\alpha$ emission
and a significant amount of continuum emission.
A comparison of the best-fit spectral models of the 
two {\sl Chandra} observations of \clover\ 
shown in Figure 3 indicates that the Fe line and 
continuum components in the 1--2~keV band
varied in flux between observations by factors of $\sim$ 2.2 and $\sim$ 1.25, respectively.
We therefore attribute most of the detected variability 
in the 1--2~keV band between the two {\sl Chandra} observations of \clover\
to variability of the Fe line. 
If the corona is compact as in the proposed outflow geometry described in \S 3.1
we would expect the direct continuum emission from this region to be sensitive to microlensing.
Assuming the microlensing interpretation, an obvious question is why the continuum in the 
1--2 keV band has not varied as much as the Fe line component in this region.
We present two possible explanations :

(a) The X-ray flux variability between the two {\sl Chandra} observations appears to be energy dependent
in the sense that the variability amplitude increases with energy.
The ratio of the fluxes of the continuum components between the two {\sl Chandra} observations in the observed-frame bands of 
0.5--1~keV, 1--2~keV, 2--3~keV and 3--4~keV are 0.83, 1.25, 1.3, and 1.65, respectively. 
This suggests that the direct continuum emission is more absorbed than the scattered continuum at observed-frame energies of $\simlt$ 2~keV and perhaps dominates over the scattered continuum emission component at observed-frame energies of 
$\simgt$ 2~keV.
The total continuum emission in the 1--2~keV range may consist mostly of scattered emission 
coming from an extended region and less sensitive to microlensing, whereas,
at energies above 2~keV the direct continuum (which originates in a compact source) might begin to dominate 
as suggested by the possible detection of broad X-ray absorption features 
at energies above 2~keV. 

(b) Another possibility is suggested by the apparent magnification of only
the blue Fe line component during the first {\sl Chandra} observation of \clover.
This apparent magnification of only the blue line peak can be explained if 
during the first {\sl Chandra} observation
the microlensing caustic was located
on the approaching side of the accretion disk and it was moving towards the receding 
side of the disk. If this were the case we would not expect the direct continuum
to be significantly magnified by the microlensing caustic during the first {\sl Chandra} observation.
This successive brightening of the double Fe line peaks has been 
predicted in simulations of microlensing of accretion disks 
by Popovic et al. (2006).

In Figure 11a we show the 1--2~keV photon flux of image A as a function of time for the four observations of \clover.
The two \xmm\ observations do not resolve image A and we therefore provide
upper limits for these epochs. 
We notice a significant decrease in the 1-2 keV photon flux of image A 
consistent with the conclusion of the microlensing event.
In Figure 11b we show the ratio of the 1--2~keV fluxes between image A and images B+C+D for
the two \chandra\ observations of \clover.
We have also overplotted with a dashed line the ratio of the $HST$ $F702W$-band 
fluxes between image
A and images B+C+D. Since the $F706W$-band flux ratio is not sensitive to microlensing
we interpret the convergence of the 1--2~keV flux ratio to the $F702W$-band flux ratio
as the end of the microlensing event in image A.
The time-scale of the decrease of the photon flux in image A
implies that the duration of the microlensing event was of the order of 2 $\times$ 10$^{3}$ days.

The sizescale of the source region being microlensed is of the order of 
$R_{\rm src}$ = $v_{\rm t}$$t_{\rm e}$, where, $t_{\rm e}$ is the 
timescale of the microlensing event in image A and 
$v_{\rm t}$ is the velocity of the caustics in the lens plane (measured in the observer's time
frame) given by equation (B9) in Kayser, Refsdal, \& Stabell (1986). 
If we assume an observer velocity of 280~km/s (this is measured 
from the CMB dipole), a lens redshift of $z$ = 1.55, lens and AGN velocities of 600~km~s$^{-1}$,
and sum the three-dimensional velocities in equation (B9) in RMS,
we find a typical velocity of a caustic of $\sim$ 180~km~s$^{-1}$ and a 
typical sizescale of the microlensed region of 
$R_{\rm src}$ $\sim$ 3.1 $\times$ 10$^{15}$~cm $\sim$ 13 $r_{\rm g}$.

In general microlensing caustic patterns may be quite complex 
and clustered depending on the relative redshifts of the lens and source and the 
density of stars along a particular line of sight.
We have assumed that the observed decrease in the magnification of image A
is the result of a single caustic crossing whereas the 
sparse sampling of two {\sl Chandra} and two {\sl XMM-Newton} observations as shown in 
Figure 11 leaves room for multiple microlensing events between the 
two {\sl Chandra} observations.
We note, however, that shorter microlensing timescales would lead to even smaller 
sizescales of the X-ray emission region.

\section{CONCLUSIONS}

Based on our spectral analysis of a 38~ks \chandra\ observation of \clover\ 
(carried out in 2000 April and presented in Chartas et al. 2004) 
we had interpreted the strong Fe K$\alpha$ emission in the spectrum of image A 
and the energy-dependent magnification of image A 
as a result of a microlensing event in this image.
We predicted that if the microlensing interpretation was correct we should expect to detect a
reduction in the continuum and Fe K$\alpha$ emission in image A
when the magnification event was complete. 

In the follow up 89~ks observation of \clover\ 
performed about five years later 
we confirm our prediction
for the microlensing interpretation. 
Specifically, the 0.2--8~keV count-rate of image A decreased by a factor of $\sim$ 2.1 
while the 0.2--8~keV count-rates of the other images remained the same within errors
between the \chandra\ observations of \clover.
The strength of the Fe~K$\alpha$ line decreased by a factor of $\sim$ 2.3 between observations.
As a result of the decrease in the flux of image A the X-ray flux ratios of the images
varied significantly between the two \chandra\ observations of \clover.
In particular, we found that the X-ray flux ratios of the 2005 \chandra\ observation are
consistent to the $HST$ WFPC2 $F702W$-band (similar to a wide $R$-band) flux ratios which are not significantly affected by microlensing.
We therefore conclude that the microlensing event in image A ended
by the 2005 \chandra\ observation.

The observed decrease of the photon flux of image A in the 1--2~keV band
and the decrease of the ratio of the 1--2~keV fluxes between image A and images (B+C+D) 
between the two \chandra\ observations of \clover\ indicates
the end of the microlensing event in image A and 
implies that the duration of the microlensing event was of the order of 2000 days.
Assuming this timescale for a caustic crossing we estimate that the sizescale of the microlensed region is of the order of 13$r_{\rm g}$.

Our analysis of the 89~ks \chandra\ observation of \clover\ revealed
several remarkable features. The spectra of the individual images show 
high energy broad absorption features
between rest frame energies of 6.4--15~keV. These features 
are especially significant in the spectra of images
C and D where they extend from $\sim$ 6.4~keV up to energies of 
9~keV and 15~keV, respectively.

If we interpret the X-ray broad absorption features as arising from absorption 
by outflowing highly ionized Fe XXV the observed maximum 
absorption energies in images C and D imply outflow velocities of 0.29$c$ and 0.67$c$,
respectively. 
The detection of X-ray broad absorption features suggests that a fraction of the detected X-ray emission must 
be direct emission from the central source that is heavily absorbed by an outflow.

The available spectra of the individual images 
cannot provide tight constraints on the properties of the 
high-energy broad absorption features. 
However, we found that fits to the spectra of images C and D with 
models that include a direct, power-law continuum
%direct power-law emission 
modified by intrinsic 
neutral absorption, a Thomson scattered power-law continuum,
saturated high-energy absorption from the outflowing material  
and a fluorescent Fe line from an accretion disk 
provide a significant improvement compared to fits with models
that only include a power-law model modified by neutral intrinsic absorption.

We investigated various plausible mechanisms to explain the spectral differences
between images. We concluded that variability of the absorber on 
time scales shorter than the time-delays 
between images in the most likely cause of these differences.
From lens models of this system (Chae et al. 1999) we expect 
that the spectrum of image D originates at an earlier rest-frame time than that of
the spectrum of image C. This ordering would imply 
a decrease in the maximum observed radial outflow velocity
within a time shorter than the time delay
between images D and C. The time-delay between 
images D and C is estimated to be of the order of  22~days.
Plausible interpretations of the decrease in the maximum absorption energy include 
a radial deceleration of the X-ray absorbing material 
or an evolution of a flow tube across our line of sight to the emission source.

The individual spectra of the images also show emission features redward of 
the rest-frame energy of 6.4~keV that resemble the 
skewed and asymmetric structure
%double-peaked pattern
occasionally detected in the spectra of Seyfert 1 galaxies.
In the combined spectrum emission line peaks 
at 5.35 $\pm$ 0.23~keV and 
$6.3_{-0.3}^{+0.6}$~keV are detected at the 
$ \simgt$  98\%  and $\simgt$ 99\% confidence levels, respectively.
The respective equivalent widths of the emission line peaks are 
$\sim500_{-200}^{+1200}$~eV and $\sim1000_{-900}^{+1200}$~eV.
In LoBAL quasars we expect that X-ray emission from the central source and 
the near side of the accretion disk to be significantly absorbed by the outflowing
wind (see Figure 8).  In \clover\ we therefore propose as plausible 
origins of the emission line peaks fluorescence from 
the far side of the accretion disk and/or from the far side of the outflowing wind.
The large equivalent widths of the line peaks can be explained by the 
significant absorption of the central continuum source by the outflowing X-ray absorbing material.
Unfortunately the present data cannot constrain the column density
of this possibly Compton-thick component.
We note that the moderate detection levels of the
double-peaked Fe K$\alpha$ line in the 89~ks observation
of \clover\ combined with the independent microlensing constraint of the sizescale of
the emitting region of $\sim$ 13 $r_{\rm g}$ and the extreme redshift of the line peak at 
5.35~keV provide a convincing case for the relativistic nature of the detected Fe K$\alpha$ line. 

We modeled the Fe~K$\alpha$ profile with an accretion disk line model
and found an acceptable fit to the spectrum.
We note, however, that due to the low S/N of the spectrum
we cannot obtain tight constraints on the parameters of such a 
disk-line model.  
Additional deeper observations are needed to 
confirm the origin of the X-ray broad absorption features, determine the cause of the
spectral differences between images, determine the origin of the Fe emission line and constrain 
the parameters of a disk-line model. 
Constraining the inclination angle of \clover\ will allow us to test the unification scheme between 
BAL and non-BAL quasars. 
A detection of a small inclination angle (i.e., $\simlt$ 50 degrees)
would challenge the validity of the orientation model of BAL quasars.
On the other hand the detection of a large inclination angle (i.e., $\simgt$ 50 degrees) would
provide supporting evidence for the unification model,
however, additional observations of other BAL quasars would be needed to reject
competing BAL models. Specifically, to reject the "youth" model that proposes LoBALs are quasars 
at an early stage in their evolution (see for example Voit, Weymann \& Korista 1993; Becker et al. 2000; 
Lacy et al. 2002)
we would need several observations of BAL quasars that show that the inclination angle 
is consistently large whereas the "youth" model
predicts a more uniform distribution of inclination angles.
A confirmation of the unification model 
is very important in cross-checking the frequency of outflows in quasars and by 
extension establishing their importance in shaping the evolution of their host galaxies
and in regulating the growth of the central black hole.

The present detection of disk like Fe~K$\alpha$ line
and highly blueshifted X-ray broad absorption features in \clover\
indicate that this is an ideal system to study quasar outflows.
The detection of the double-peaked Fe emission line is possible partly due the the 
significant absorption of the central source resulting in an increased equivalent width
and the large flux magnification from lensing expected to be between 20--40.

\acknowledgments
We acknowledge financial support from NASA grants NAS8-03060 and NAG5-10817.
We thank Chris Kochanek and Luca Popovic for fruitful 
discussions related to the origin of the detected Fe emission lines in H~1413+117 
and microlensing of quasars. ME acknowledges partial support from the Theoretical Astrophysics Visitors' Fund at Northwestern University and thanks the members of the group for their warm hospitality. 

\small

\clearpage

\normalsize

\beginrefer

\refer Arnaud, K.~A.\ 1996, ASP 
Conf.~Ser.~101: Astronomical Data Analysis Software and Systems V, 5, 17 \\

\refer Becker, R.~H., White, 
R.~L., Gregg, M.~D., Brotherton, M.~S., Laurent-Muehleisen, S.~A., \& Arav, 
N.\ 2000, ApJ, 538, 72 \\

\refer {Braito}, V., {Della Ceca}, R., {Piconcelli}, E., et al.,
2004, \aap, 420, 79 \\

\refer Chae \& Turnshek\ 1999, ApJ, 514 587 \\

\refer Chae, K.-H., Turnshek, 
D.~A., Schulte-Ladbeck, R.~E., Rao, S.~M., \& Lupie, O.~L.\ 2001, \apj, 
561, 653 \\

\refer Chartas, G., Brandt, W.~N., Gallagher, S.~C., \& Garmire, G.~P.\ 2002, \apj, 579, 169 \\

\refer Chartas, G., Brandt, W.~N., Gallagher, S.~C., \& Proga, D. \ 2007, accepted in AJ, astro-ph/0701104 \\

\refer Chartas, G., Brandt, W.~N., \& Gallagher, S.~C.\ 2003, ApJ, 595, 85 \\

\refer Chartas, G., Eracleous, M., Agol, E., \& Gallagher, S.~C.\ 2004, \apj, 606, 78 \\

\refer Chelouche, D. 2003, ApJ, 579, 169 \\

\refer Collin, S., Coup{\'e}, 
S., Dumont, A.-M., Petrucci, P.-O., \& R{\'o}{\.z}a{\'n}ska, A.\ 2003, 
\aap, 400, 437 \\

\refer Czerny, B., 
R{\'o}{\.z}a{\'n}ska, A., Dov{\v c}iak, M., Karas, V., \& Dumont, A.-M.\ 
2004, \aap, 420, 1 \\

\refer Dai, X., Chartas, G., Eracleous, M., \& Garmire, G.~P.\ 2004, \apj, 605, 45\\

\refer Dai, X., Kochanek, C. S., Chartas, G., \& Mathur, S. \ 2006, \apj, 637, 53\\

\refer Dickey, J.~M., \& Lockman, F.~J.\ 1990, \araa, 28, 215 \\

\refer Dov{\v c}iak, M., 
Karas, V., \& Yaqoob, T.\ 2004, \apjs, 153, 205 \\

\refer Elvis, M.\ 2000, \apj, 545, 63 \\

\refer Fabian, A.~C., Rees,  M.~J., Stella, L., \& White, N.~E.\ 1989, \mnras, 238, 729 \\

\refer Falco, E. E., Impey, C. D., Kochanek, C. S. et al. 1999, ApJ, 523, 617 \\

\refer Gabel, J.~R., et al.\ 2003, \apj, 595, 120 \\

\refer Gallagher, S.~C., Brandt, W.~N., Chartas, G., \& Garmire, G.~P.\ 2002,  \apj, 567, 37 \\

\refer Gallagher, S.~C., 
Schmidt, G.~D., Smith, P.~S., Brandt, W.~N., Chartas, G., Hylton, S., 
Hines, D.~C., \& Brotherton, M.~S.\ 2005, \apj, 633, 71 \\

\refer Garmire, G.~P., Bautz, M.~W., Ford, P.~G., Nousek, J.~A., \& Ricker, G.~R.\ 2003, \procspie, 4851, 28 \\

\refer George, I.~M., \& 
Fabian, A.~C.\ 1991, \mnras, 249, 352 \\

\refer Ghisellini, G., 
Haardt, F., \& Matt, G.\ 2004, \aap, 413, 535 \\

\refer Goodrich, R.~W., \& Miller, J.~S.\ 1995, \apjl, 448, L73 \\

\refer Granato, G.~L., De 
Zotti, G., Silva, L., Bressan, A., \& Danese, L.\ 2004, \apj, 600, 580 \\

\refer Green, P. J.\ 2006, \apj, 664, 733 \\

\refer Green, P.~J., Aldcroft, 
T.~L., Mathur, S., Wilkes, B.~J., \& Elvis, M.\ 2001, \apj, 558, 109\\

\refer Haardt, F., \& Maraschi, L.\ 1991, \apjl, 380, L51 \\

\refer Hopkins, P.~F., 
Hernquist, L., Cox, T.~J., Di Matteo, T., Martini, P., Robertson, B., \& 
Springel, V.\ 2005, \apj, 630, 705 \\

\refer Hopkins, P.~F., 
Hernquist, L., Cox, T.~J., Di Matteo, T., Robertson, B., \& Springel, V.\ 
2006, \apjs, 163, 1 \\

\refer Jansen, F., Lumb, D., Altieri, B., et al. 2001, A\&A, 365, L1\\

\refer Jim{\'e}nez-Bail{\'o}n, E., Piconcelli, E., Guainazzi, M., Schartel, N., 
Rodr{\'{\i}}guez-Pascual, P.~M., \& Santos-Lle{\'o}, M.\ 2005, \aap, 435, 
449 \\

\refer Krolik, J.~H.\ 1998, Active 
Galactic Nuclei: From the Central Black Hole to the Galactic Environment, 
by J.H.~Krolik.~Princeton: Princeton University Press, 1998.\\

\refer Lacy, M., Gregg, M., 
Becker, R.~H., White, R.~L., Glikman, E., Helfand, D., \& Winn, J.~N.\ 
2002, \aj, 123, 2925 \\

\refer Laor, A.\ 1991, \apj, 376, 90 \\

%\refer Maloney, P.~R., \& Reynolds, C.~S.\ 2000, \apjl, 545, L23 \\

\refer Martocchia, A., 
Matt, G., \& Karas, V.\ 2002, \aap, 383, L23 \\

\refer Merloni, A.\ 2003, \mnras, 341, 1051 \\

\refer Miniutti, G., \& Fabian, A.~C.\ 2006, \mnras, 366, 115 \\

\refer Mori, K., Tsunemi, H., Miyata, E., Baluta, C., Burrows, D. N.,
Garmire, G. P., \& Chartas, G. 2001, in ASP Conf. Ser. 251, New Century
of X-Ray Astronomy, ed. H. Inoue \& H. Kunieda (San Francisco: ASP), 576 \\

\refer Nandra, K., George, 
I.~M., Mushotzky, R.~F., Turner, T.~J., \& Yaqoob, T.\ 1997, \apjl, 488, 
L91 \\

\refer Oshima, T., Mitsuda, K., Fujimoto, R., Iyomoto, N., Futamoto, K., Hattori, M., Ota, N.,
Mori, K., Ikebe, Y., Miralles, J.~M., \&  Kneib, J.-P. \ 2001, \apjl, 563, L103 \\

\refer Popovi{\'c}, L.~{\v 
C}., Jovanovi{\'c}, P., Mediavilla, E., Zakharov, A.~F., Abajas, C., 
Mu{\~n}oz, J.~A., \& Chartas, G.\ 2006, \apj, 637, 620 \\

\refer Porquet, D., Reeves, 
J.~N., O'Brien, P., \& Brinkmann, W.\ 2004, \aap, 422, 85 \\

\refer Proga, D.\ 2005, \apjl, 630, L9\\

\refer Proga, D., Stone, J.~M., \& Kallman, T.~R.\ 2000, \apj, 543, 686 \\

\refer Reynolds, C.~S., \& 
Nowak, M.~A.\ 2003, \physrep, 377, 389 \\

\refer Schartel, N., 
Rodr{\'{\i}}guez-Pascual, P.~M., Santos-Lle{\'o}, M., Clavel, J., 
Guainazzi, M., Jim{\'e}nez-Bail{\'o}n, E., \& Piconcelli, E.\ 2005, \aap, 
433, 455 \\

\refer Springel, V., Di Matteo, T., \& Hernquist, L.\ 2005, \apjl, 620, L79 \\

\refer Scannapieco, E., \& Oh, S.~P.\ 2004, \apj, 608, 62 \\

\refer {Str{\" u}der}, L., {Briel}, U., {Dennerl}, K., et al.,
 \ 2001, \aap, 365, L18 \\

%\refer Tanaka, Y., Nandra, K., Fabian, A.~C. et al., 1995, \nat, 375, 659 \\  
	
\refer Tsunemi, H., Mori, K., 
Miyata, E., Baluta, C., Burrows, D.~N., Garmire, G.~P., \& Chartas, G.\ 
2001, \apj, 554, 496 \\

\refer {Turner}, M.~J.~L., {Abbey}, A., {Arnaud}, M., 
\ 2001, \aap, 365, L27 \\

%\refer Turner, T.~J., \& Kraemer, S.~B.\ 2003, \apj, 598, 916 \\

%\refer Turner, T.~J., Miller, L., George, I.~M., \& Reeves, J.~N.\ 2006, \aap, 445, 59\\

\refer Turnshek, D. A., Lupie, O. L., Rao, S. M., Espey, B. R.,
\& Sirola, C. J. 1997, \apj, 485, 100 \\

\refer Voit, G.~M., Weymann, 
R.~J., \& Korista, K.~T.\ 1993, \apj, 413, 95 \\

\refer Wilms, J., Allen, A., \& McCray, R.\ 2000, \apj, 542, 914 \\

\refer Wise, M. W., Davis, J. E., Huenemoerder, Houck, J. C., Dewey, D.
Flanagan, K. A., and Baluta, C. 1997,
{\it The MARX 3.0 User Guide, CXC Internal Document}
available at http://space.mit.edu/ASC/MARX/ \\

\endrefer

\clearpage
\scriptsize
\begin{center}
\begin{tabular}{lcccc}
\multicolumn{5}{c}{TABLE 1}\\
\multicolumn{5}{c}{Log of Observations of the LoBAL Quasar \clover} \\
 & & & & \\ \hline\hline
                  &                      &                     & Effective  &      \\
Observation & Observatory    &  Observation  &   Exposure Time\tablenotemark{a}  & $R_{src}$\tablenotemark{b}    \\
Date           &                      &  ID                &    (ks)     &     \\
&    &    &    &     \\
\hline
%&    &    &    &     \\
%2000 April 19     & {\it Chandra}          & 930           & 38.19 &  38.2  &  8.2 $\pm$ 0.5 $\times$ 10$^{-3}$   \\
%2005 March 30   & {\it Chandra}          & 5645         & 88.86 &   88.9  & 5.6 $\pm$ 0.3 $\times$ 10$^{-3}$  \\
%2001 July 29         & {\it XMM-Newton} & 0112250301 &  19.56        & 19.2~ks    &  1.06 $\pm$ 0.1 $\times$ 10$^{-2}$\\
%2002 August 2         & {\it XMM-Newton} &0112251301  &  29.64      & 23.5~ks  &  0.94 $\pm$ 0.1 $\times$ 10$^{-2}$ \\
2000 April 19     & {\it Chandra}          & 930           &   38.2  &  8.2 $\pm$ 0.5 $\times$ 10$^{-3}$   \\
2001 July 29         & {\it XMM-Newton} & 0112250301 &   19.2    &  1.1 $\pm$ 0.1 $\times$ 10$^{-2}$\\
2002 August 2         & {\it XMM-Newton} &0112251301  &  23.5  &  0.9 $\pm$ 0.1 $\times$ 10$^{-2}$ \\
2005 March 30   & {\it Chandra}          & 5645         &    88.9  & 5.6 $\pm$ 0.3 $\times$ 10$^{-3}$  \\
%&    &    &    &   &  \\
\hline \hline
\end{tabular}
\end{center}
${}^{a}${The effective exposure time is the time remaining after the application of good time-interval (GTI)
tables to remove portions of the observation that were severely contaminated by background flares.}\\
${}^{b}${Background-subtracted source count rate including events with energies within the 0.2--10~keV band.
The source count rates and effective exposure times for the \xmm\ observations refer to those obtained with the
EPIC PN instrument.
See \S 2 for details on source and background extraction regions used for estimating $R_{src}$.}\\

\clearpage
\scriptsize
\begin{center}
\begin{tabular}{cccccc}
\multicolumn{6}{c}{TABLE 2}\\
\multicolumn{6}{c}{RESULTS FROM FITS TO THE {\it Chandra} SPECTRA OF INDIVIDUAL IMAGES} \\
 & & & & &\\ \hline\hline
\multicolumn{1}{c} {Model$^{a}$} &
\multicolumn{1}{c} {Parameter$^{b}$} &
\multicolumn{1}{c} {Values For} &
\multicolumn{1}{c} {Values For} &
\multicolumn{1}{c} {Values For} &
\multicolumn{1}{c} {Values For} \\

        &           & Image A$^{c}$  & Image B$^{c}$&                             Image C$^{c}$                 &       Image D$^{c}$                         \\ \hline
                             &           &   &&                                              &       \\    
 1 &$\Gamma$     &  0.62 $\pm$ 0.21 & 0.38 $\pm$ 0.21 &   0.55 $\pm$ 0.23   &  0.50 $\pm$ 0.26  \\
 &$C-statistic/nbins$ & 433/783 & 496.3/783 & 406.1/783 & 364/783\\
  &$\chi^2/{\nu}$ & 35.4/25 & 30.9/28 & 39.3/22 & 27.5/16 \\
&$P(\chi^2/{\nu})$$^{d}$ & 8.1~$\times$~10$^{-2}$  &0.3 &  1.3~$\times$~10$^{-2}$  & 3.9~$\times$~10$^{-2}$ \\

    &           &   &&                                              &       \\

2 &$\Gamma$                                          & 1.7$_{-0.7}^{+0.4}$    & 0.93$_{-0.42}^{+0.33}$   &   1.9$_{-0.4}^{+0.7}$     &1.64$_{-0.40}^{+0.28}$   \\
&  $N_{\rm H}$ (10$^{22}$~cm$^{-2}$ )& 19$_{-13}^{+9}$ & 9$_{-5}^{+7}$ & 29$_{-10}^{+21}$  & 24$_{-10}^{+9}$\\
 &$C-statistic/nbins$                                & 412.7/783                            & 487/783                     & 380/783                        & 352/783\\
  &$\chi^2/{\nu}$ & 20.9/24 & 23.7/27 & 22.9/21 & 17.6/15\\
&$P(\chi^2/{\nu})$$^{d}$ & 0.64  & 0.65 & 0.35  & 0.28 \\

    &           &   &&                                              &       \\

3 &$\Gamma$                                     & --      &-- &      1.5$_{-0.5}^{+0.6}$(68\%) &1.7$^{fp}$\\
&$N_{\rm H}$ (10$^{22}$~cm$^{-2}$ ) & --&--&         25$_{-14}^{+21}$ (68\%)         & 120$_{-50}^{+100}$ (68\%)  \\
&$E_{\rm}$ (keV)                                    & --&--&       5.6 $\pm$ 0.1 (68\%)                               & 5.6$_{-0.8}^{+0.6}$ (68\%)  \\
&$\sigma_{\rm}$ (keV)                                &-- &--&    0.24 $_{-0.16}^{+0.15}$ (68\%)             &1.7 $_{-0.4}^{+0.5}$ (68\%)\\                                           
&$E_{\rm BAL}$ (keV)                                &--&--&   8.5 $\pm$ 0.2 (68\%)                  &13.9 $\pm$ 0.2 (68\%)\\
&$w_{\rm BAL}$ (keV)                            & --&--& 0.5$_{-0.5}^{+0.3}$ (68\%)            &2.2 $\pm$ 0.4 (68\%)  \\
%&$EW_{\rm BAL}$                                  && & 2.42$_{}^{+3.34}$~keV \\
%&$\chi^2/{\nu}$                                    &&& 19.6/22 &\\
%&$P(\chi^2/{\nu})$$^{e}$                     && & 0.6  &\\
 &$C-statistic/nbins$                                &--&--&       365/783           &  330/783\\
   &$\chi^2/{\nu}$                            & &                 & 12.0/16 & 8.3/11\\
&$P(\chi^2/{\nu})$$^{d}$               &  &                 & 0.75  & 0.68 \\
    &           &   &&                                              &       \\

4 &$\Gamma$                                     & --      &-- &      2.1 $\pm$ 0.6 (68\%) &1.6$_{-0.6}^{+1.1}$(68\%) \\
&$N_{\rm H}$ (10$^{22}$~cm$^{-2}$ ) & --&--&         55$_{-23}^{+43}$ (68\%)         & 79$_{-46}^{+77}$ (68\%)  \\
&$E_{\rm}$ (keV)                                    & --&--&       6.4$^{fp}$                               & 6.4$^{fp}$ \\
%&$EW_{\rm} (keV)$                                & &&     $_{-0}^{+}$  &\\
&$r_{\rm in}$ ($r_{\rm g}$)                       & --&--&     6 ${}^{fp}$                              & 6 ${}^{fp}$  \\
&$r_{\rm out}$ ($r_{\rm g}$)                      & --&--&    1000 ${}^{fp}$                         & 1000 ${}^{fp}$\\
&$q$                                                          &--&--&     4$^{fp}$                                    & 4$^{fp}$\\
&$i$ (degrees)                                             &-- &--&    21 $\pm$ 7 (68\%)             &27 $_{-10}^{+7}$ (68\%)\\                                           
&$E_{\rm BAL}$ (keV)                                &--&--&   8.5 $\pm$ 0.2 (68\%)                  &13.9 $\pm$ 0.2 (68\%)\\
&$w_{\rm BAL}$ (keV)                            & --&--& 0.5$_{-0.5}^{+0.3}$ (68\%)            & 2.3 $\pm$ 0.4 (68\%)  \\
%&$EW_{\rm BAL}$                                  && & 2.42$_{}^{+3.34}$~keV \\
%&$\chi^2/{\nu}$                                    &&& 19.6/22 &\\
%&$P(\chi^2/{\nu})$$^{e}$                     && & 0.6  &\\
 &$C-statistic/nbins$                                &--&--& 365/783           &  327/783\\
   &$\chi^2/{\nu}$                            & &                 & 14.8/16 & 7.8/10\\
&$P(\chi^2/{\nu})$$^{d}$               &  &                 & 0.54  & 0.64 \\
    &           &   &&                                              &       \\                        

5 &$\Gamma$                                     &    --   &-- &      2.9$_{-0.7}^{+1.2}$(68\%) &1.7 $\pm$ 0.9 (68\%)\\
&$N_{\rm H}$ (10$^{22}$~cm$^{-2}$ ) &-- &--&         110$_{-35}^{+69}$ (68\%)         & 78$_{-56}^{+66}$ (68\%)  \\
&$E_{\rm}$ (keV)                                    & --&--&       6.4$^{fp}$                               & 6.4$^{fp}$ \\
%&$EW_{\rm} (keV)$                                & &&     $_{-0}^{+}$  &\\
&$r_{\rm in}$ ($r_{\rm g}$)                       &-- &--&     1.47 ${}^{fp}$                              & 1.67 ${}^{fp}$  \\
&$r_{\rm out}$ ($r_{\rm g}$)                      &-- &--&    1.52 ${}^{fp}$                         & 1.69 ${}^{fp}$\\
&$q$                                                          &--&--&     4$^{fp}$                                    & 4$^{fp}$\\
&$i$ (degrees)                                             & --&--&    69 $\pm$ 1 (68\%)             &66 $_{-10}^{+7}$ (68\%)\\                                           
&$E_{\rm BAL}$ (keV)                                &--&--&   8.6$_{-0.2}^{+0.1}$ (68\%)                  &14.0 $\pm$ 0.2 (68\%)\\
&$w_{\rm BAL}$ (keV)                            &-- &--& 0.5 $\pm$ 0.3 (68\%)            & 2.3 $\pm$ 0.4 (68\%)  \\
%&$EW_{\rm BAL}$                                  && & 2.42$_{}^{+3.34}$~keV \\
%&$\chi^2/{\nu}$                                    &&& 19.6/22 &\\
%&$P(\chi^2/{\nu})$$^{e}$                     && & 0.6  &\\
 &$C-statistic/nbins$                                &--&--& 362/783           &  328/783\\
    &           &   &&                                              &       \\

  & &  &&                   &                                          \\
\hline \hline
\end{tabular}
\end{center}
\noindent
${}^{a}$ Model 1 consists of a power law 
Model 2 consists of a power law and neutral absorption at the source. 
Model 3 consists of emission of a power-law modified by intrinsic neutral absorption at the source, 
saturated high-energy absorption, and a Gaussian emission line.
Model 4  consists of direct emission of a power-law modified by intrinsic neutral absorption at the source, 
scattered emission of a power law assuming simple Thomson scattering, 
saturated high-energy absorption, and
a fluorescent Fe line from an accretion disk around a Schwarzschild black hole,
where the Fe line model is based on Fabian et al. (1989).
Model 5  consists of direct emission of a power-law modified by intrinsic neutral absorption at the source, 
scattered emission of a power law assuming simple Thomson scattering, 
saturated high-energy absorption, and
a fluorescent Fe line from an accretion disk around a Kerr black hole,
where the Fe line model is based on Laor et al. (1991).
All model fits include the Galactic absorption toward the source (Dickey \& Lockman, 1990).\\
${}^{b}$All absorption-line parameters are calculated for the rest frame.\\
${}^{c}$All errors are for 90\% confidence unless mentioned otherwise with all
parameters taken to be of interest except absolute normalization.\\
${}^{d}$$P(\chi^2/{\nu})$ is the probability of exceeding $\chi^{2}$ for ${\nu}$ degrees of freedom
if the model provides an acceptable description of the data. \\
%${}^{e}$ $f_{c}$ is the covering fraction. \\

\clearpage
\scriptsize
\begin{center}
\begin{tabular}{ccccc}
\multicolumn{2}{c}{TABLE 3}\\
\multicolumn{2}{c}{RESULTS FROM FITS TO THE COMBINED SPECTRUM OF ALL {\it Chandra} IMAGES OF THE 2005 OBSERVATION} \\
 &  \\ \hline\hline
\multicolumn{1}{c} {Parameter$^{a}$} &
\multicolumn{1}{c} {Values For} \\
                & Combined Spectrum ${}^{b}$ 
                &  \\ \hline
\multicolumn{2}{c} {Model 1: Two Gaussian Emission Lines and One Gaussian Absorption Line${}^{c}$} 
                &                         \\                        
$\Gamma$                                            &  0.7 $\pm$ 0.2  \\
$N_{\rm H}$ (10$^{22}$~cm$^{-2}$ ) & 9$_{-4}^{+5}$  \\
$E_{\rm red} (keV)$                                     &  5.35 $\pm$ 0.23 \\
$\sigma_{\rm red} (keV)$                             & 0.15$_{-0.15}^{+0.24}$(68\%) \\
$EW_{\rm red}(keV)$                                   & 0.5$_{-0.2}^{+1.2}$ \\
$E_{\rm blue} (keV)$                                     &  6.3$_{-0.3}^{+0.6}$ \\
$\sigma_{\rm blue} (keV)$                             & 0.5$_{-0.3}^{+0.4}$ (68\%) \\
$EW_{\rm blue} (keV)$                                   & 1.0$_{-0.9}^{+1.2}$ \\
$E_{\rm BAL} (keV)$                                     &  14 $\pm$ 2 \\
$\sigma_{\rm BAL} (keV)$                             & 2.6$_{-0.7}^{+1.6}$ (68\%) \\
$EW_{\rm BAL} (keV)$                                   & 2.4$_{-1.6}^{+3.3}$ \\
$\chi^2/{\nu}$                                    & 17.6/20 \\
$P(\chi^2/{\nu})$$^{f}$                      & 0.6  \\
  &                                                    \\
  \multicolumn{2}{c} {Model 2:  Disk-Line and One Gaussian Absorption Line${}^{d,g}$} \\
$\Gamma$                                            &  0.74$_{-0.22}^{+0.23}$  \\
$N_{\rm H}$ (10$^{22}$~cm$^{-2}$ ) & 10$_{-4}^{+6}$  \\
$E_{\rm} (keV)$                                     &  6.4$^{fp}$  \\
$EW_{\rm} (keV)$                                   & $1.2 \pm 0.5$  \\
$r_{\rm in}$ ($r_{\rm g}$)                       & 15.7 ${}^{up}$ \\
$r_{\rm out}$ ($r_{\rm g}$)                      & 15.8 ${}^{up}$ \\
$q$                                                          &4$^{fp}$ \\
$i$ (degrees)                                             & 26 $_{-6}^{+13}$ \\                                           
$E_{\rm BAL}$                                     &  13$_{-4}^{+9}$~keV \\
$\sigma_{\rm BAL}$                             & 3$_{-1}^{+4}$~keV (68\%) \\
%$EW_{\rm BAL}$                                   & 2.42$_{}^{+3.34}$~keV \\
$\chi^2/{\nu}$                                    & 19.6/22 \\
$P(\chi^2/{\nu})$$^{f}$                      & 0.6  \\
  &                                                    \\
   \multicolumn{2}{c} {Model 3:  Kerr Disk-Line and One Gaussian Absorption Line${}^{e,g}$} \\
$\Gamma$                                            &  0.8$_{-0.3}^{+0.2}$  \\
$N_{\rm H}$ (10$^{22}$~cm$^{-2}$ ) & 11$_{-6}^{+5}$  \\
$E_{\rm} (keV)$                                     &  6.4$^{fp}$  \\
$EW_{\rm} (keV)$                                   & 1.6$_{-0.9}^{+0.9}$  \\
$r_{\rm in}$ ($r_{\rm g}$)                       & 1.46 ${}^{up}$ \\
$r_{\rm out}$ ($r_{\rm g}$)                      & 1.83 ${}^{up}$ \\
$q$                                                          &4$^{fp}$ \\
$i$ (degrees)                                             & 70$_{-8}^{+12}$ \\                                           
$E_{\rm BAL}$                                     &  13$_{-2}^{+3}$~keV \\
$\sigma_{\rm BAL}$                             & 2.6$_{-0.9}^{+1.0}$~keV (68\%) \\
%$EW_{\rm BAL}$                                   & 2.42$_{}^{+3.34}$~keV \\
$\chi^2/{\nu}$                                    & 22/22 \\
$P(\chi^2/{\nu})$$^{f}$                      & 0.45  \\
  &                                                    \\

%  \multicolumn{2}{c} {Model 3: Disk-Line, Gaussian Emission Line, Gaussian Absorption Line${}^{c}$}\\ 
%$\Gamma$                                            &  0.69$_{-0.22}^{+0.23}$  \\
%$N_{\rm H}$ (10$^{22}$~cm$^{-2}$ ) & 8.9$_{-3.7}^{+5.3}$  \\
%$E_{\rm}$                                     &  6.32$_{-0.31}^{+0.56}$~keV \\
%$\sigma_{\rm}$                             & 0.49$_{-0.30}^{+0.42}$~keV (68\%) \\
%$EW_{\rm}$                                   & 1043$_{-893}^{+1193}$~keV \\
%$E_{\rm blue}$                                     &  6.32$_{-0.31}^{+0.56}$~keV \\
%$\sigma_{\rm blue}$                             & 0.49$_{-0.30}^{+0.42}$~keV (68\%) \\
%$EW_{\rm blue}$                                   & 1043$_{-893}^{+1193}$~keV \\
%$E_{\rm BAL}$                                     &  13.47$_{-2.21}^{+1.87}$~keV \\
%$\sigma_{\rm BAL}$                             & 2.80$_{-1.06}^{+5.58}$~keV (68\%) \\
%$EW_{\rm BAL}$                                   & 2.42$_{-1.60}^{+3.34}$~keV \\
%$\chi^2/{\nu}$                                    & 17.6/20 \\
%$P(\chi^2/{\nu})$$^{f}$                      & 0.6  \\

\hline \hline
\end{tabular}
\end{center}
\noindent
${}^{a}$All absorption-line parameters are calculated for the rest frame.\\
${}^{b}$All errors are for 90\% confidence unless mentioned otherwise with all
parameters taken to be of interest except absolute normalization.\\
${}^{c}$ Model 1 consists of a power law, neutral absorption at the source, two Gaussian emission lines at the source and one Gaussian absorption line at the source.
In XSPEC notation this is written as:
wabs*zwabs*(zgauss + zgauss + zgauss + pow).\\
${}^{d}$ Model 2 consists of a power-law modified by intrinsic 
neutral absorption, a Gaussian absorption line,
and a fluorescent Fe line from an accretion disk around a Schwarzschild black hole.
In XSPEC notation this is written as:
wabs*zwabs*(diskline + zgauss + pow).\\
${}^{e}$ Model 3 consists of a power-law modified by intrinsic 
neutral absorption, a Gaussian absorption line,
and a fluorescent Fe line from an accretion disk around a Kerr black hole.
In XSPEC notation this is written as :
wabs*zwabs*(laor + zgauss + pow).\\
${}^{f}$$P(\chi^2/{\nu})$ is the probability of exceeding $\chi^{2}$ for ${\nu}$ degrees of freedom
if the model provides an acceptable description of the data. \\
${}^{g}$  ${fp}$ indicates a parameter is fixed and ${up}$ indicates a parameter
unconstrained.
% is not well constrained.
All model fits include the Galactic absorption toward the source (Dickey \& Lockman, 1990).

\clearpage
\scriptsize
\begin{center}
\begin{tabular}{clccc}
\multicolumn{5}{c}{TABLE 4}\\
\multicolumn{5}{c}{RESULTS FROM FITS TO THE {\it XMM-NEWTON} SPECTRA OF \clover} \\
 & & &  &\\ \hline\hline
\multicolumn{1}{c} {Fit} &
\multicolumn{1}{c} {Model$^{a}$} &
\multicolumn{1}{c} {Parameter} &
\multicolumn{1}{c} {\xmm\ Value$^{b}$} &
\multicolumn{1}{c} {\xmm\ Value$^{b}$} \\
  &                                         &   & 2001 July 29                  &     2002 August 2                           \\
    &                                         &   &                   &                       \\

 $ 1 $& PL and neutral absorption        &$\Gamma$ &  1.7$_{-0.3}^{+0.3}$  &      1.1$_{-0.3}^{+0.3}$     \\
  &  at source                                  & $N_{\rm H}$ (10$^{22}$~cm$^{-2}$ )& 25$_{-8}^{+9}$  &  19$_{-8}^{+11}$\\
  &                               & $\chi^2/{\nu}$              & (47.5)/43)                    & (57.0)/(43) \\
  &                               & $P(\chi^2/{\nu})$$^{c}$           & 0.3&  0.18    \\
  &                                                             &           &                   & \\
  &                               &                             &                                  &        \\
  &                               &                   &                                         & \\

${2}$   & PL and neutral absorption &$\Gamma$ & 1.4$_{-0.3}^{+0.3}$ & 0.9$_{-0.3}^{+0.3}$    \\
  & and emission line at source.       & $N_{\rm H}$ (10$^{22}$~cm$^{-2}$ ) & 19$_{-7}^{+9}$  &   14$_{-8}^{+9}$     \\
  &                               & E$_{\rm emis}$    & 5.8$_{-2.5}^{+0.2}$~keV                 & $6.2_{-0.1}^{+0.3}$~keV \\
    &                               & $\sigma_{\rm emis}$    & 0.35$_{-0.12}^{+0.25}$~keV                 & $ < 0.65$~keV  \\
  &                               & $\chi^2/{\nu}$    &   (41.5)/(38)                                       &    (47.6)/(38)  \\
  &                               &$P(\chi^2/{\nu})$$^{c}$ &   0.32                        &  0.14 \\
  &                               &                   &                                          & \\
  &                               &                   &                                          & \\
\hline \hline
\end{tabular}
\end{center}
${}^{a}$All model fits include Galactic absorption of $N_{\rm H}$ = 1.82 $\times$ 10$^{20}$~cm$^{-2}$ toward the source (Dickey \& Lockman, 1990).
All fits to the \xmm\ spectra are performed on the PN, MOS1 and MOS2 spectra. \\
${}^{b}$All errors are for 68\% confidence with all
parameters taken to be of interest except absolute normalization.\\
${}^{c}$$P(\chi^2/{\nu})$ is the probability of exceeding $\chi^{2}$ for ${\nu}$ degrees of freedom
if the model provides an acceptable description of the data.\\

\clearpage
\begin{figure*}
\centerline{\includegraphics[width=14cm]{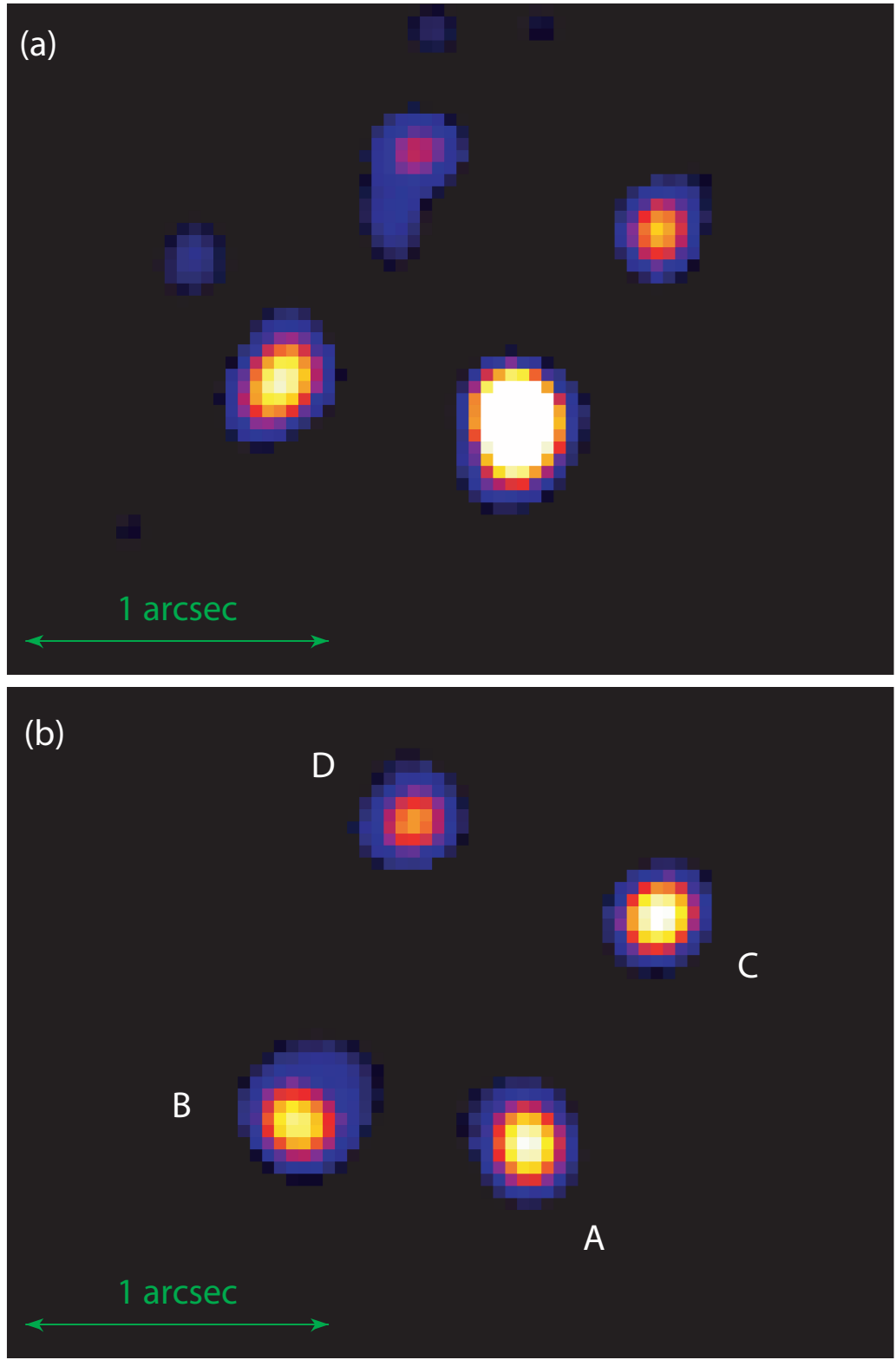}}
\caption{\small  The Lucy-Richardson deconvolved images in the 0.2--8~keV bandpass of the
38~ks (panel a) and 89~ks (panel  b) \chandra\ observations of \clover\, respectively. 
The brightness scale of the images represents count rate.
%The images have been normalized by the exposure times.
The significant decrease of the X-ray flux of only image A is interpreted as the result of a 
microlensing event in image A that peaked near the 38~ks observation. The X-ray flux ratios of the images during 
the 89~ks observation are consistent with the {\it HST} $F702W$-Band flux ratios. The images are displayed 
with a linear brightness scale. East is to the left and north is up.
\label{fig1.eps}}
\end{figure*}

\clearpage
\begin{figure*}
\centerline{\includegraphics[width=9cm]{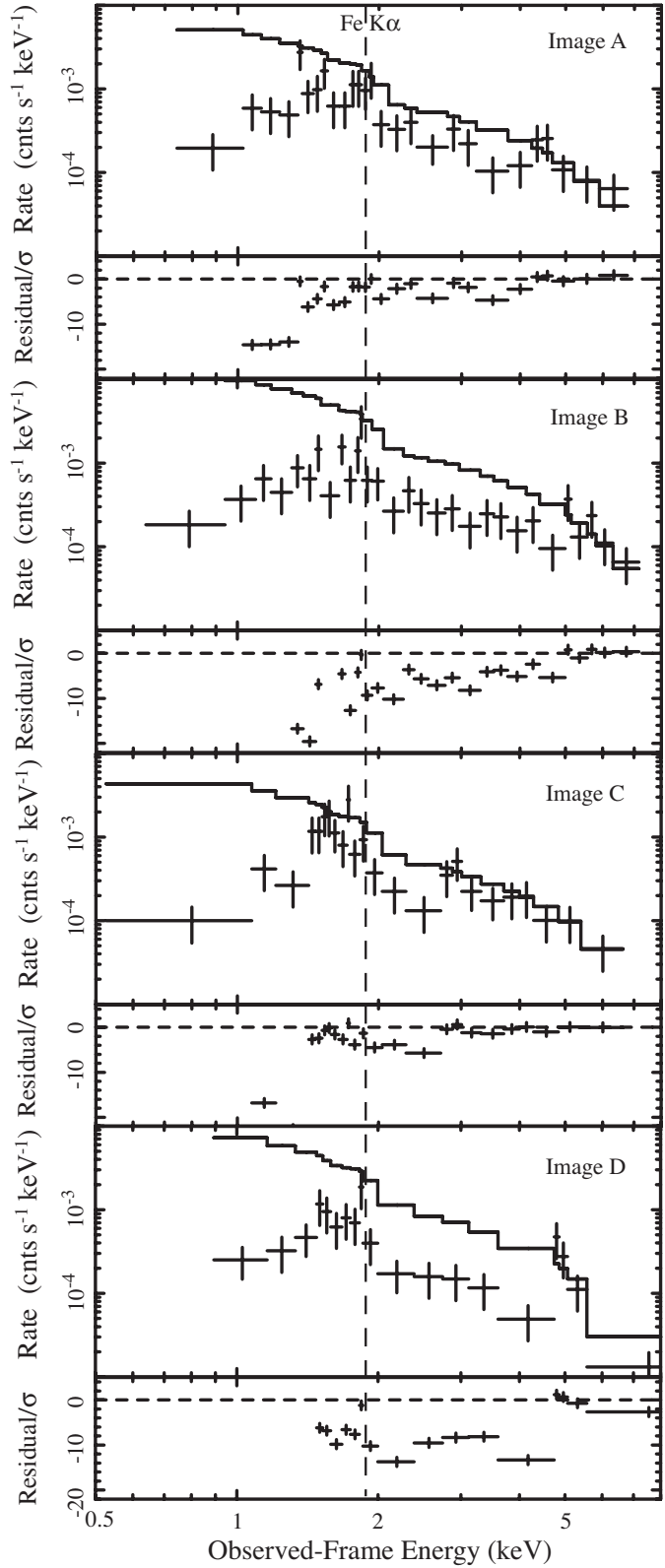}}
\caption{\small  The spectra of images A, B, C, and D 
from the 89~ks {\it Chandra} observation of \clover\ fitted with
a power-law model and Galactic absorption.
The model was fitted to points in the following energy range:
4.5--8~keV for image A, 5--8~keV for image B, 3--8~keV for image C and 4--8~keV for image D and then extrapolated to lower energies.
The dashed vertical line indicates the location of the resdshifted
Fe K$\alpha$ emission line. 
\label{fig2.eps}}
\end{figure*}

\clearpage
\begin{figure*}
\centerline{\includegraphics[width=14cm]{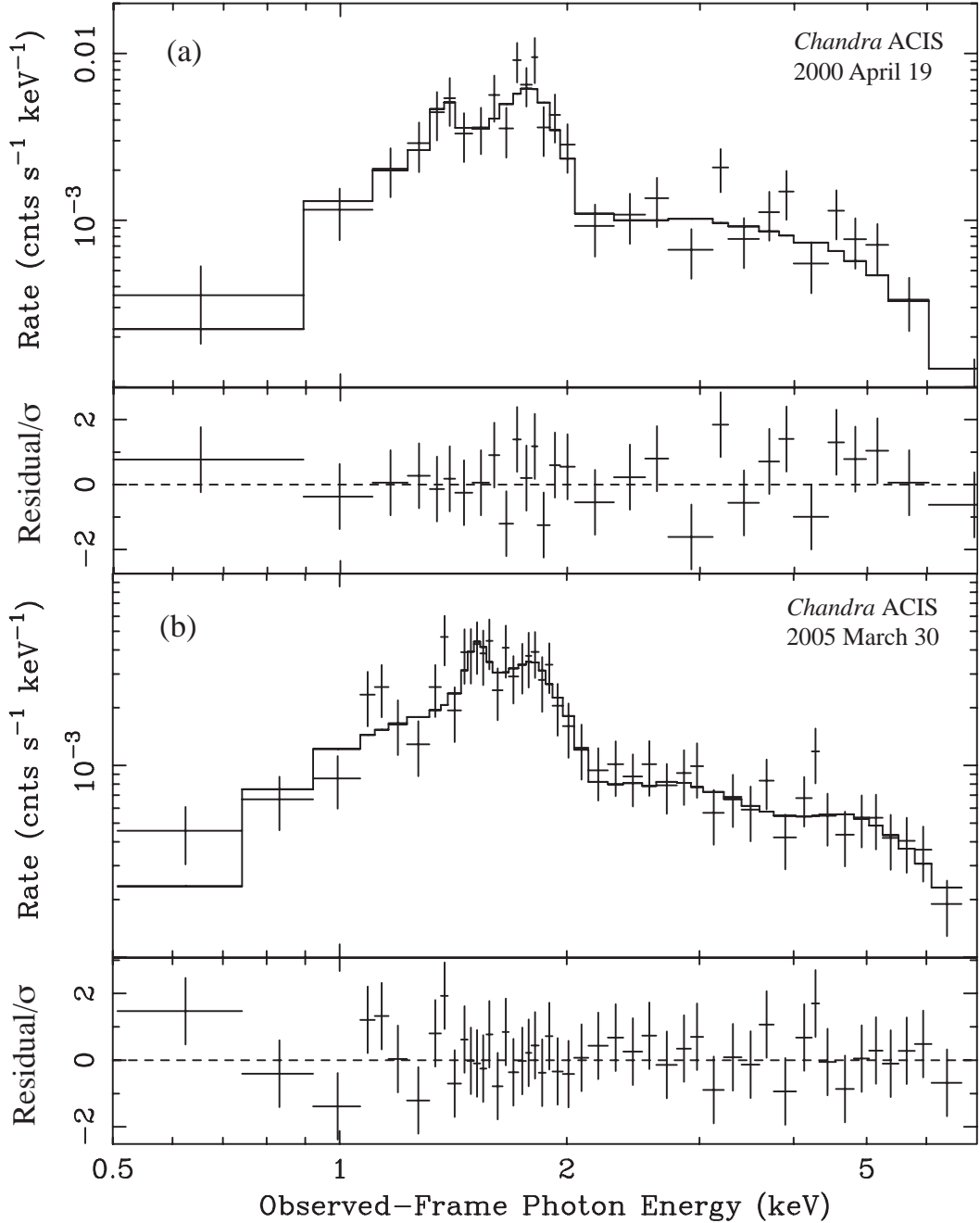}}
\caption{\small  
The spectra of combined images A, B, C, and D of \clover\
from (a) the 38~ks {\it Chandra} observation
and (b) the 89~ks {\it Chandra} observation. 
The spectra were fitted with
a model that consists of a power law, neutral absorption at the redshift of the source,
a Gaussian absorption line at the redshift of the 
source and two Gaussian emission lines at the redshift of the source.
\label{fig3.eps}}
\end{figure*}

\clearpage
\begin{figure*}
\centerline{\includegraphics[width=14cm]{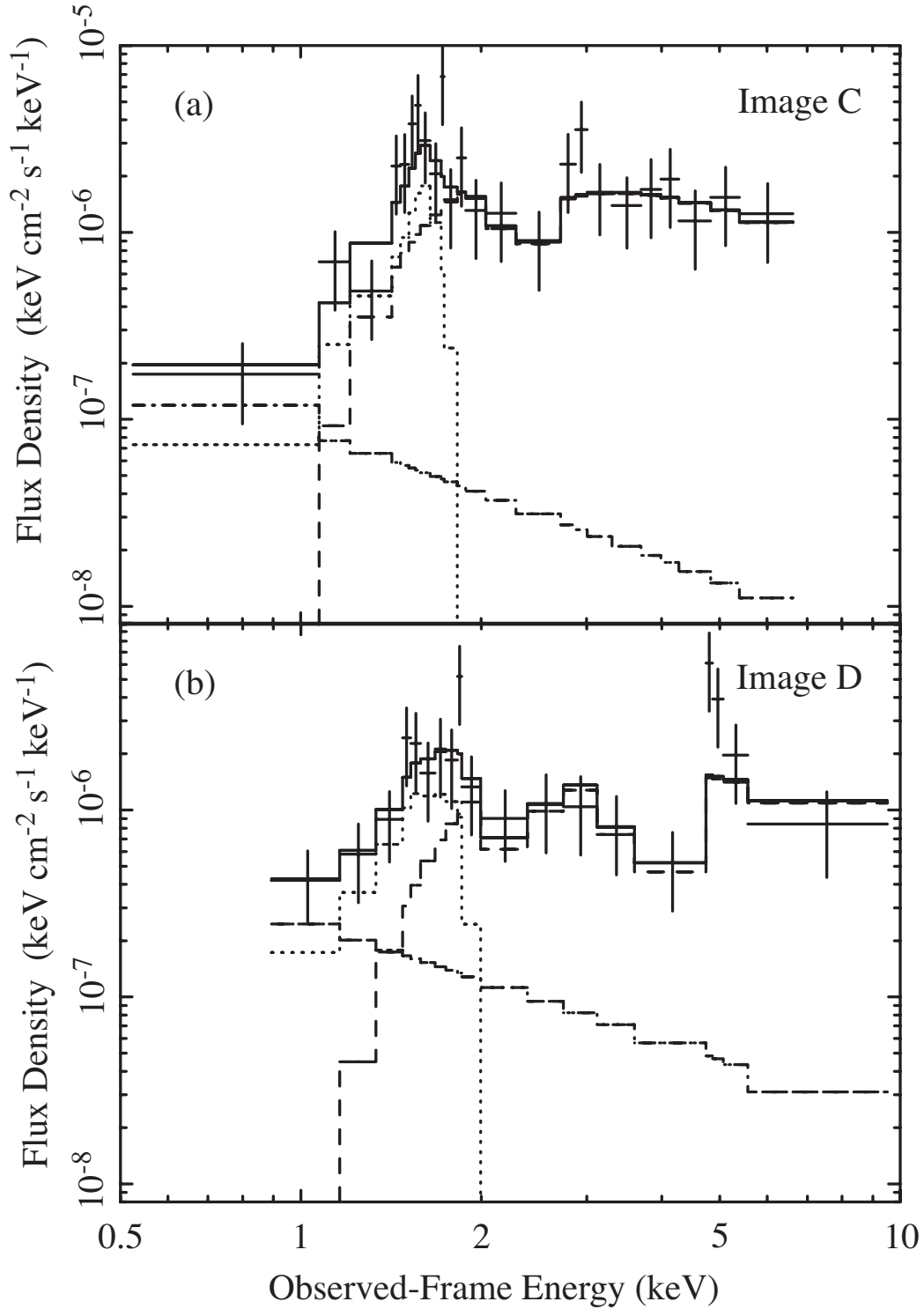}}
\caption{\small  The unfolded 2005 \chandra\ spectra of images C (panel a) and D 
(panel b) of \clover\ plotted with the best-fit model that consists of 
the following components:
Galactic absorption, direct emission of a power-law modified by intrinsic neutral absorption 
(dashed line), scattered emission of a power law assuming simple Thomson scattering 
(dot-dashed line), a fluorescent Fe line from an accretion disk around a black hole
where the Fe line model is based on Laor's (1991) calculation 
applicable to a Kerr black hole (dotted line),
%and includes GR effects (dotted line),
and saturated high-energy absorption (see text for more details; Model 5 of  Table 2).
We emphasize that the present data cannot adequately constrain such a complex model,
however, the main purpose of these fits is to demonstrate that such a 
%plausible 
model is consistent with the data (model 1 from Table 3).
\label{fig4.eps}}
\end{figure*}

\clearpage
\begin{figure*}
\centerline{\includegraphics[width=14cm]{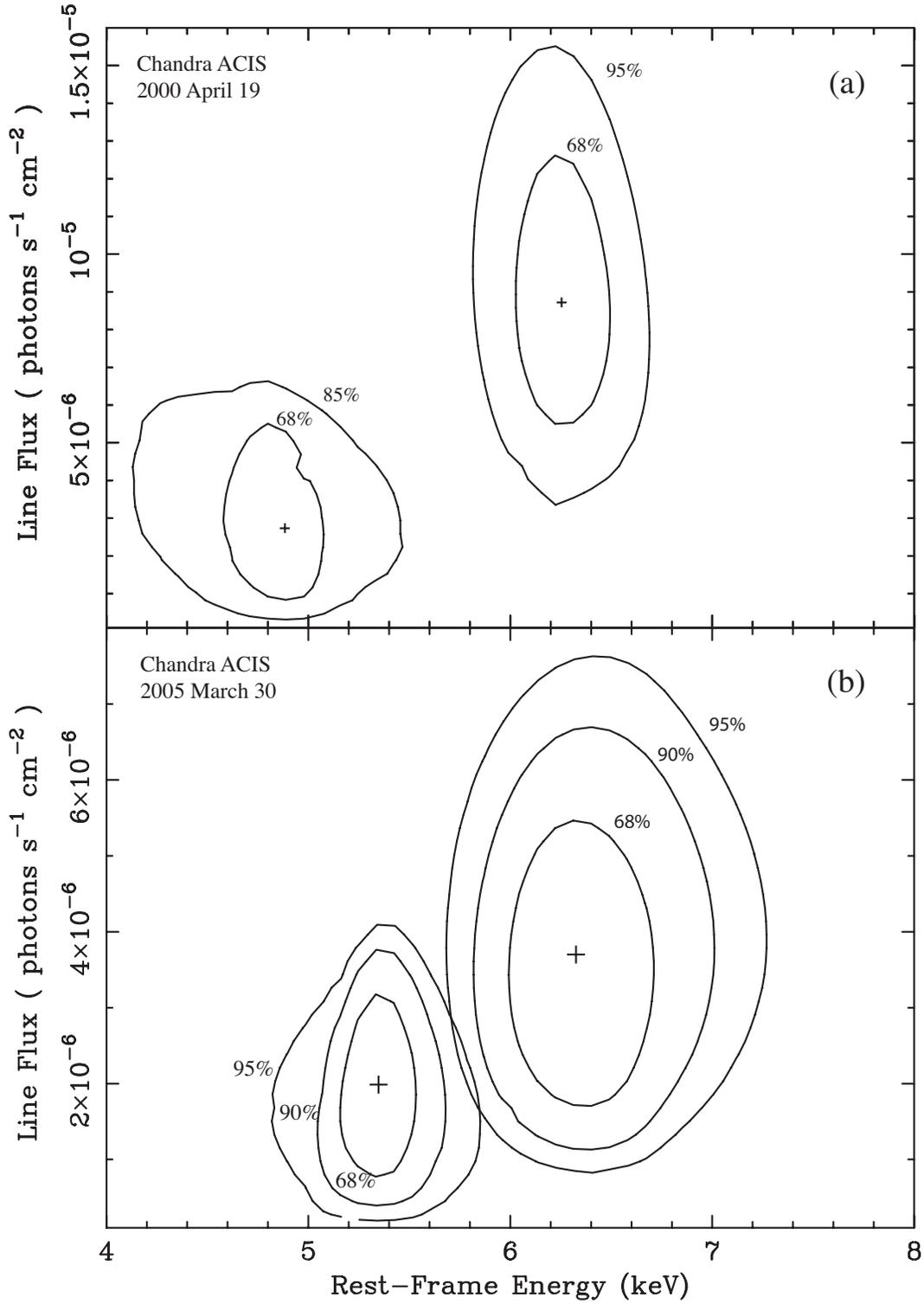}}
\caption{\small Confidence contours between 
the flux and energy of the two Fe K emission line peaks detected in the combined 
spectra of all images 
of (a) the 38~ks \chandra\ observation and (b)
the 89~ks \chandra\ observation of \clover.  
\label{fig5.eps}}
\end{figure*}

\clearpage
\begin{figure*}
\centerline{\includegraphics[width=14cm]{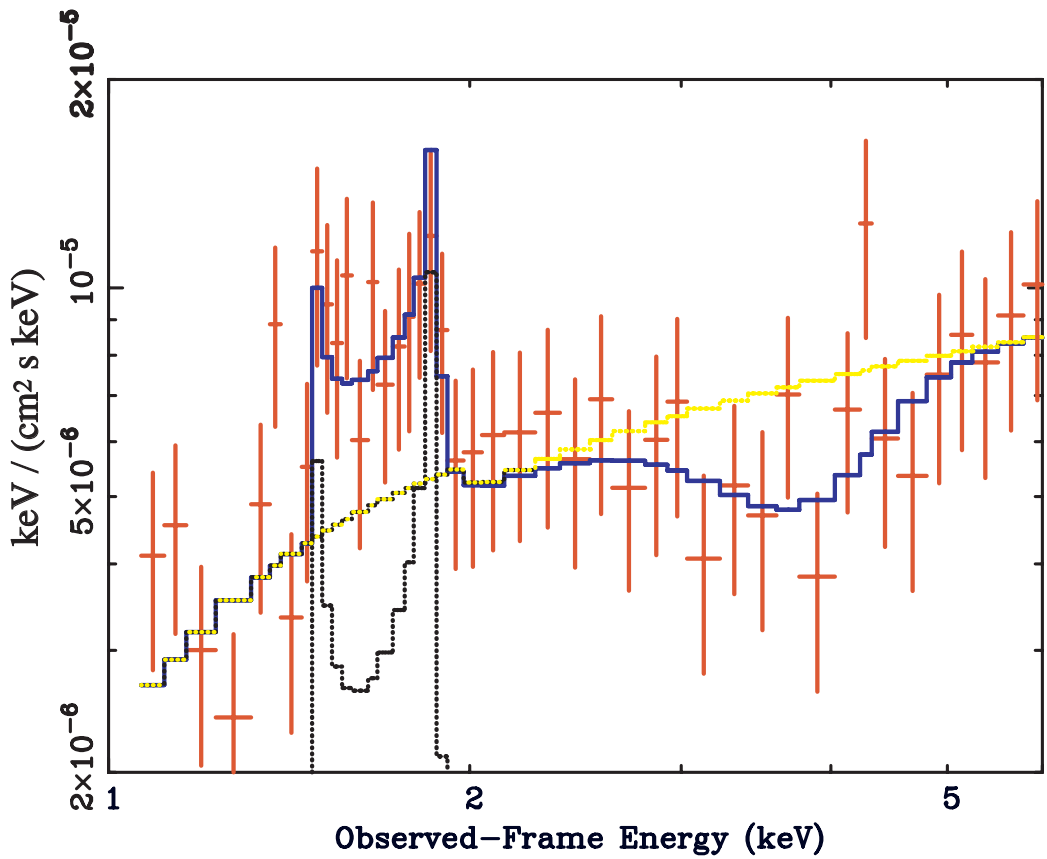}}
\caption{\small  The unfolded 2005 \chandra\ spectrum of \clover\ plotted with the best-fit model that consists of 
Galactic and intrinsic neutral absorption,
a power law, a broad absorption line and a fluorescent Fe line from an accretion disk around a black hole. The Fe line model
is from Fabian et al. (1989) and 
is applicable to a Schwarzschild black hole
%includes GR and Doppler effects 
(see model 2 in Table 3).
\label{fig6.eps}}
\end{figure*}

\clearpage
\begin{figure*}
\centerline{\includegraphics[width=14cm]{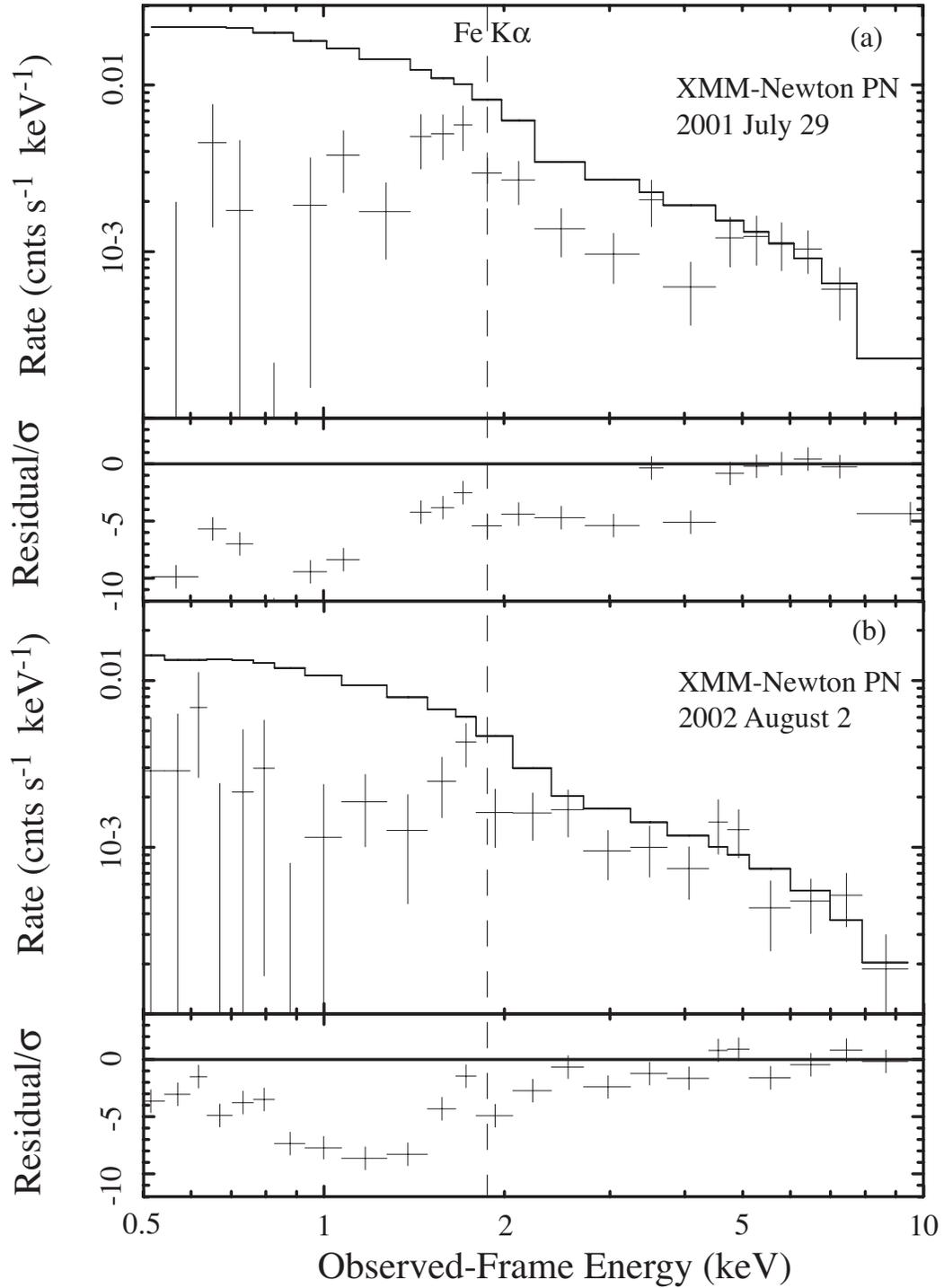}}
\caption{\small  The combined \xmm\ PN spectra of images A, B, C, and D of \clover\
for the (a) the 2001 July 29 19.2~ks observation and
(b) the 2002 August 2 23.5~ks observation.
The 4.5--10~keV spectra were fitted with a power-law model modified 
by Galactic absorption. This model was then extrapolated to lower energies.
The dashed vertical line indicates the location of the redshifted Fe~K$\alpha$
emission line. 
\label{fig7.eps}}
\end{figure*}

\clearpage
\begin{figure*}
\centerline{\includegraphics[width=14cm]{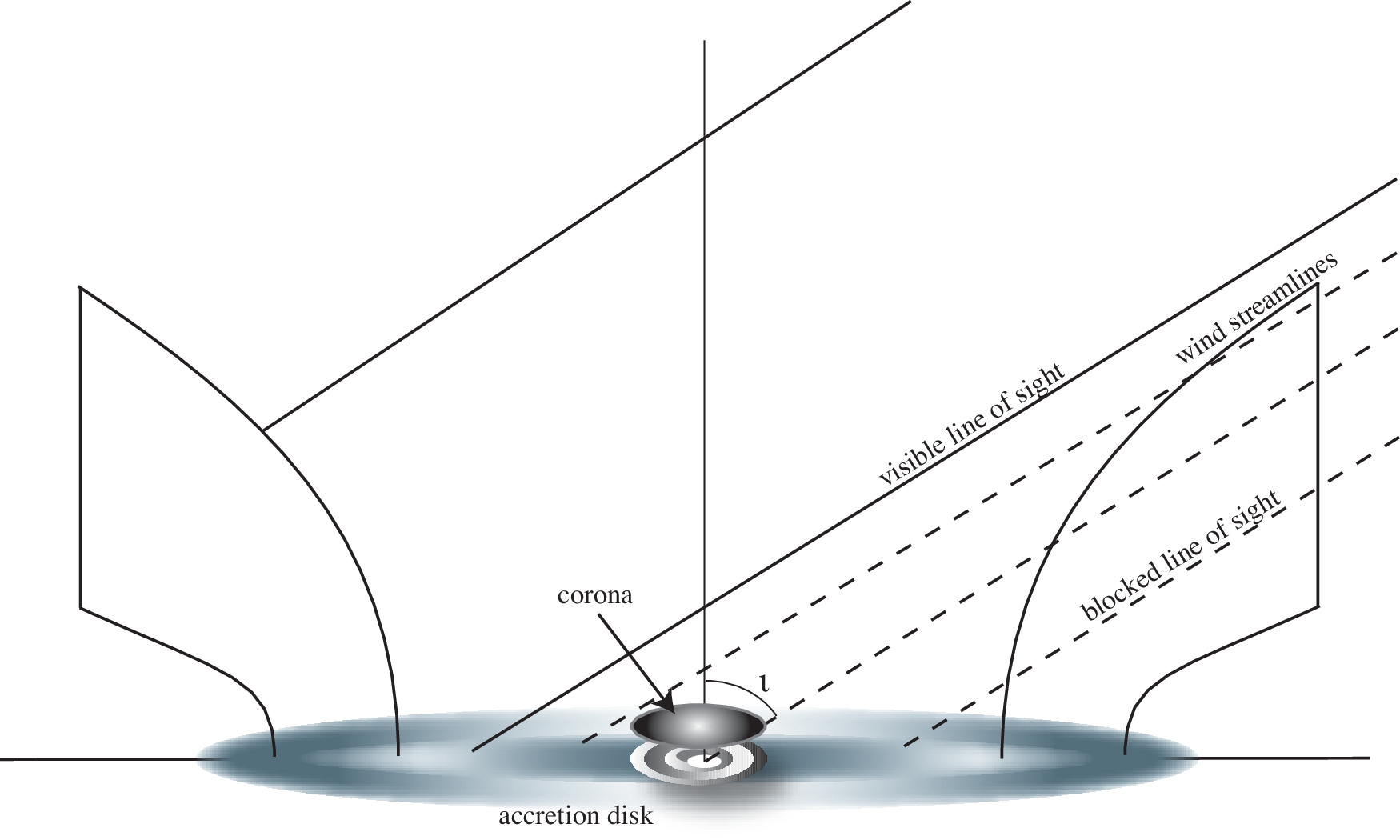}}
\caption{\small
Schematic diagram of a proposed geometry for the 
accretion disk and associated outflow in \clover. 
%outflow and accretion disk of \clover. 
X-ray emission from the near side of the 
accretion disk and the central continuum source is blocked by the 
Compton thick absorbing wind. Scattered and fluorescent emission
from the far side of the accretion disk and outflow may reach the observer.
Light rays that originate near the black hole will be slightly bent due to GR effects.
\label{fig8.eps}}
\end{figure*}

\clearpage
\begin{figure*}
\centerline{\includegraphics[width=14cm]{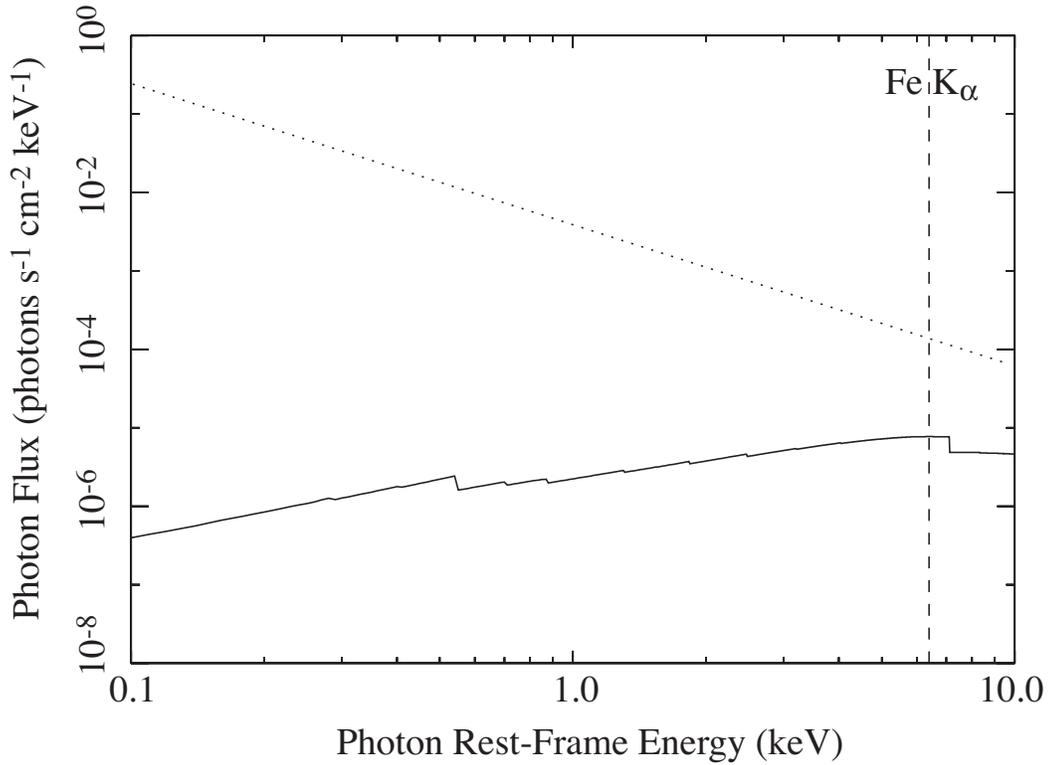}}
\caption{\small The dotted line shows the photon flux density of the central source 
which we assume to scale as $F_{\rm cs}(E)$ $\propto$ $E^{-\Gamma}$.
The solid line shows the scattered emission spectrum, $F_{\rm scat}(E)$,
assuming Thomson scattering of the central source spectrum by a scatterer that subtends 
a solid angle of $\Omega_{\rm scat}$ to the central source.
We have assumed  $\Gamma$ = 1.8,  $\Omega_{\rm scat}$ = 0.15 and 
standard solar abundance values after Wilms, Allen, \& McCray (2000).
The vertical dashed line indicates the location of the Fe~K$\alpha$ line.
The discontinuities in the scattered spectrum are photoelectric absorption edges.
They result from the fact that the scattered photons travel through 
a significant column of the (mostly neutral) scattering medium.
\label{fig9.eps}}
\end{figure*}

\clearpage
\begin{figure*}
\centerline{\includegraphics[width=14cm]{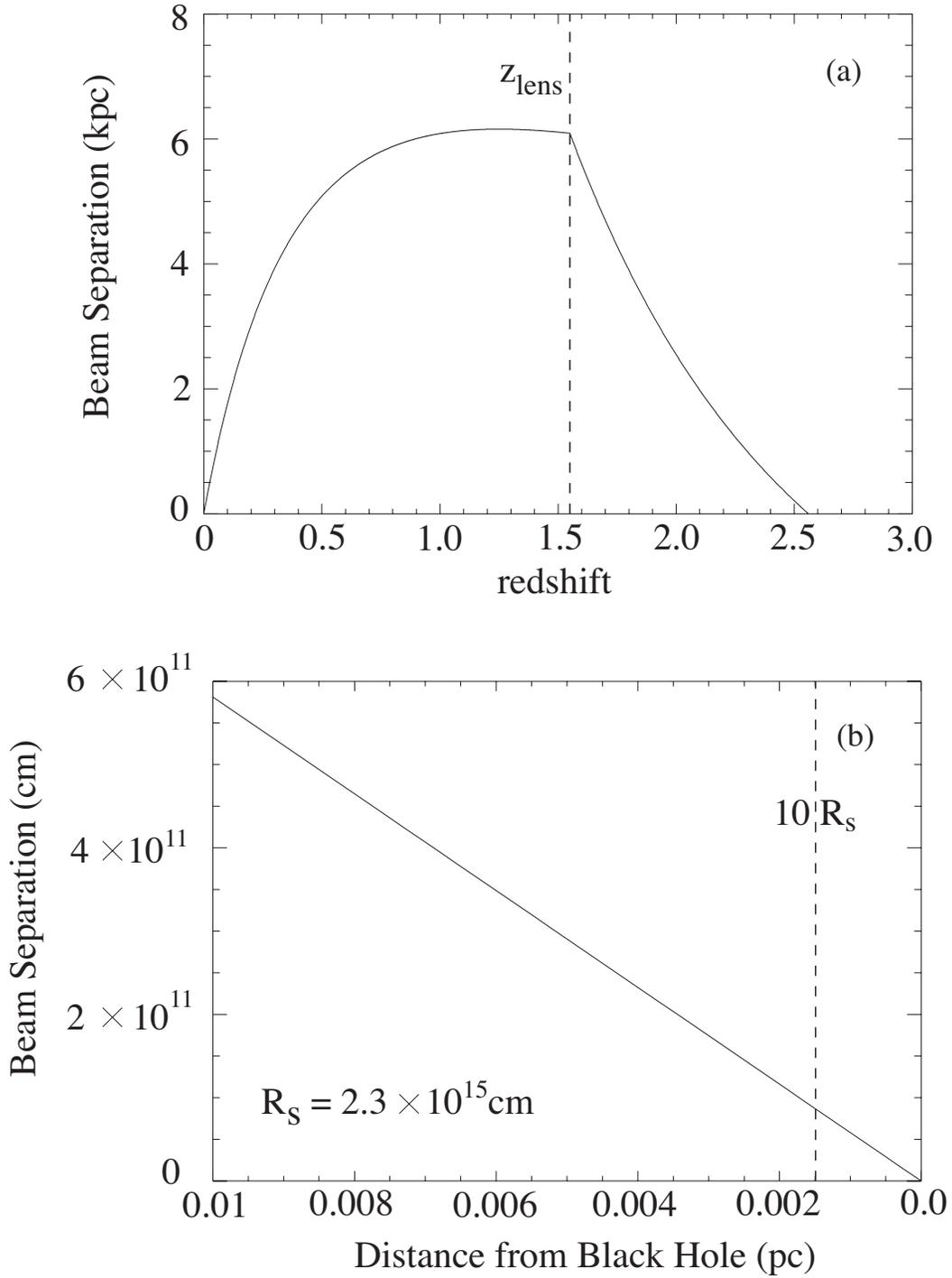}}
\caption{\small 
The beam separation as a function of (a) redshift 
assuming an observed image separation of 1~arcsec 
and lens and quasar redshifts of 1.55 and 2.56, respectively,
and (b) distance from the black hole assuming a lens redshift of $ z $= 1.55.
The dashed vertical line marks a distance of 10 Schwartzschild radii for a
black hole mass of \hbox{$M_{\rm BH}$ = 1.5 $\times$ 10$^{9}$ M$_{\odot}$}. %(see \S3.3 of the text).
\label{fig10.eps}}
\end{figure*}

\clearpage
\begin{figure*}
\centerline{\includegraphics[width=14cm]{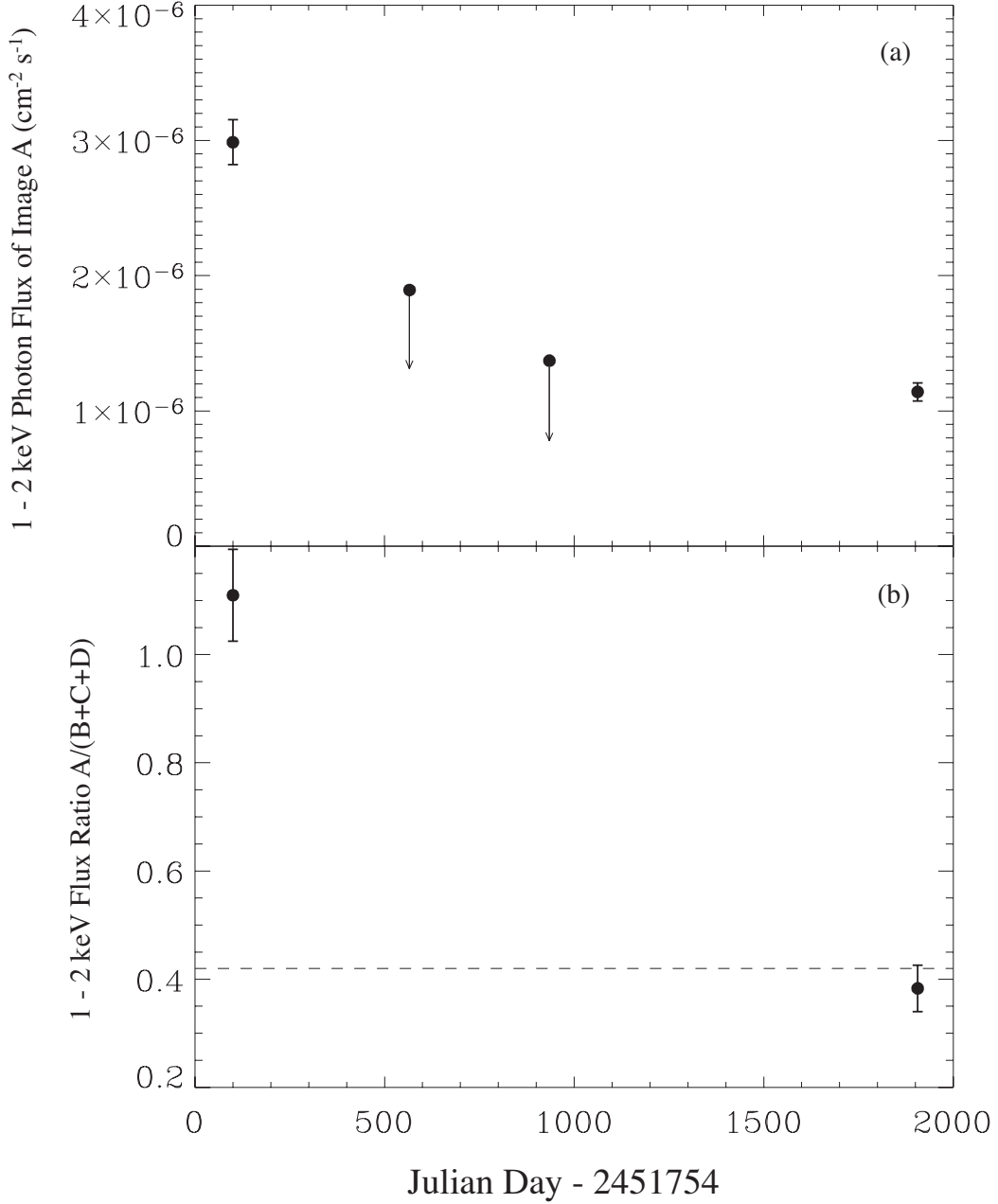}}
\caption{\small
(a) The 1--2~keV photon flux of image A for the four observations of \clover.
The two \xmm\ observations do not resolve image A but provide
upper limits for these epochs. 
(b) The ratio of the 1--2~keV fluxes between image A and images B+C+D for
the two \chandra\ observations of \clover.
We have also overplotted with a dashed line the ratio of the $HST$ $F702W$-band fluxes between image
A and images B+C+D. Since the $F702W$-band flux ratio is less sensitive to microlensing
we interpret the convergence of the 1--2~keV flux ratio to the $F702W$-band flux ratio
as the end of the microlensing event in image A.
 \label{fig11.eps}}
\end{figure*}

\clearpage
\begin{figure*}
\centerline{\includegraphics[width=14cm]{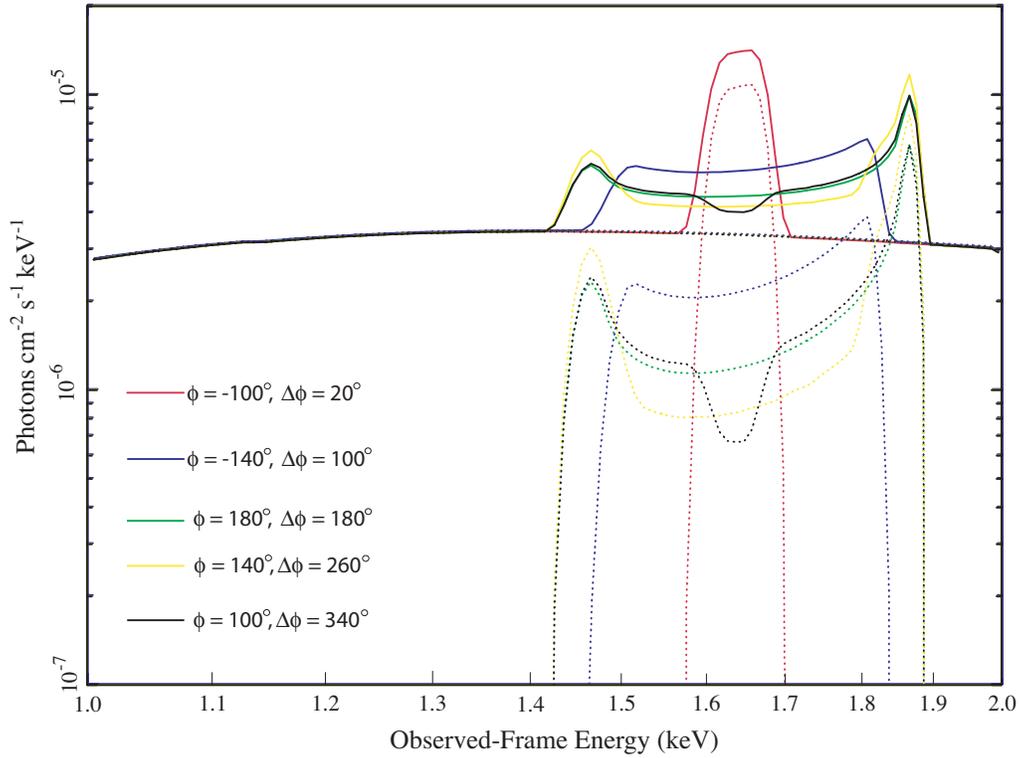}}
\caption{\small Fe~K$\alpha$ line profiles originating from five different azimuthal segments
of the disk assuming that portions of the disk are obscured by the outflowing
wind as illustrated in Figure 8. $kyr1line$ (Dovciak et al. 2004) was used to model the relativistic line from an accretion disc around a Kerr black hole in the case of non-axisymmetric disk emission. The inclination angle, and inner and outer radii of the disk
emission were fixed to the values of 30~degrees, 14~$r_{\rm g}$ and 
17~$r_{\rm g}$, respectively, found from fits to the \chandra\ spectrum of \clover\ using the axisymmetric model $kyrline$.  
%Models with 180 $ > $ $\phi$ $ > $ 90 are consistent with the data. 
Models with $-140$ degrees $ <  \phi <  -90 $ degrees, where substantially 
more than half of the disk emission is obscured by the wind, are not consistent with the data.
$\phi$ (units of degrees) is the lower azimuth of non-zero disk emissivity
and $\Delta\phi$ (units of degrees) is the span of the disk sector with non-zero disc emissivity.
 \label{fig12.eps}}
\end{figure*}

\end{document}